\documentclass[reprint,twocolumn,aps,prd,nofootinbib,superscriptaddress]{revtex4-2}

\usepackage{amsmath}	
\usepackage{amssymb}
\usepackage{mathrsfs}
\usepackage[dvipsnames]{xcolor}
\usepackage{graphicx}   
\usepackage{hyperref}
\usepackage{times}

\newcommand*{\ns}{_{\mathrm{NS}}}

\newcommand{\eg}{e.g.,~}
\newcommand{\ie}{i.e.,~}

\hypersetup{colorlinks=true,        
linkcolor=blue,                     
citecolor=cyan,                     
}

\begin{document}


\title{Numerical modelling of bulk viscosity in neutron stars}

\author{Michail Chabanov}
\email[]{micsma@rit.edu}
\affiliation{Institut f\"ur Theoretische Physik, Goethe Universit\"at,
  Max-von-Laue-Str. 1, 60438 Frankfurt am Main, Germany}
\affiliation{Center for Computational Relativity and Gravitation \& School of Mathematical
Sciences, Rochester Institute of Technology, 85 Lomb Memorial Drive, Rochester,
New York 14623, USA}

\author{Luciano Rezzolla}
\affiliation{Institut f\"ur Theoretische Physik, Goethe Universit\"at,
Max-von-Laue-Str. 1,
60438 Frankfurt am Main, Germany}
\affiliation{Frankfurt Institute for Advanced Studies,
Ruth-Moufang-Str. 1,
60438 Frankfurt am Main, Germany}
\affiliation{School of Mathematics, Trinity College,
Dublin 2, Ireland}

\date{\today}

\begin{abstract}
The early post-merger phase of a binary neutron-star coalescence is
shaped by characteristic rotational velocities as well as violent density
oscillations and offers the possibility to constrain the properties of
neutron star matter by observing the gravitational wave emission. One
possibility to do so is the investigation of gravitational wave damping
through the bulk viscosity which originates from violations of weak
chemical equilibrium. Motivated by these prospects, we present a
comprehensive report about the implementation of the self-consistent and
second-order formulation of the equations of relativistic hydrodynamics
for dissipative fluids proposed by M\"uller, Israel and Stewart.
Furthermore, we report on the results of two test problems, namely the
viscous damping of linear density oscillations of isolated nonrotating
neutron stars and the viscous migration test, both of which confirm our
implementation and can be used for future code tests. Finally, we present
fully general-relativistic simulations of viscous binary neutron-star
mergers. We explore the structural and thermal properties of binary
neutron-star mergers with a constant bulk-viscosity prescription and
investigate the impact of bulk viscosity on dynamical mass ejection. We
find that inverse Reynolds numbers $\sim 1\%$ can be achieved for the
highest employed viscosity thereby suppressing the dynamically ejected
mass by a factor of $\sim 5$ compared to the inviscid case.
\end{abstract}


\maketitle

\section{Introduction}

Binary neutron star mergers (BNSs) are extremely violent and nonlinear
phenomena in which all of the four fundamental interactions play an
important role. This makes them ideal multi-messenger astronomical
observables which harbour the potential to place constraints on gravity
or the strong interaction of dense matter in regimes otherwise not
accessible by terrestrial experiments. For example, the first
multi-messenger detection of a BNS merger,
GW170817~\cite{Abbott2017_etal, Drout2017, Cowperthwaite2017c}, provided
constraints on the properties of isolated nonrotating neutron stars, \eg
their maximum mass $M_{_{\mathrm{TOV}}}$ or their distribution in
radii~\cite{Annala2017, Bauswein2017b, Margalit2017, Radice2017b,
  Rezzolla2017, Ruiz2017, Most2018, Shibata2019, Annala2019,
  Dietrich2020, Nathanail2021}, which can be directly used to narrow the
band of allowed equations of state (EOS) of cold nuclear matter. The
acronym TOV stands for Tolman-Oppenheimer-Volkoff. Many of these results
stem solely from the inspiral gravitational-wave (GW) signal, the
post-merger signal, however, will offer even more information about the
EOS at extreme densities ~\cite{Stergioulas2011b, Bauswein2012a,
  Takami2014, Bernuzzi2015a, Rezzolla2016, Breschi2022a}, in particular
when the possible appearance of a phase transition to quark matter is
invoked~\cite{Most2018b, Most2019c, Weih:2019xvw,
  Tootle2022,Bauswein2019, Blacker2020, Liebling2021, Prakash:2021wpz,
  Fujimoto:2022c, Ujevic2023}.

Another promising avenue to infer the EOS of dense matter through the
post-merger GW signal constitutes possible imprints of microphysical
transport effects, such as bulk viscosity, which lead to damped GW
emission. Bulk viscosity can emerge effectively through violations of
weak chemical equilibrium and results from the out-of-equilibrium
dynamics that accompanies the post-merger remnant in the first few
milliseconds since its formation~\cite{Alford2010, Alford2018,
Alford:2018lhf, Alford2021a, Hammond:2021vtv, Celora2022, Most2022_a}
(see~\cite{Ghosh2023,Ripley2023} for some recent work on the inspiral).

By considering URCA processes involving \textit{npe} or \textit{npe}$\mu$
matter several studies estimated post-merger $\mathrm{kHz}$-density
oscillations to be damped over a timescale $\lesssim 100\,\mathrm{ms}$
after the merger~\cite{Baiotti2016, Alford2020,
Alford2022a,Hammond2023a}. In addition, it was shown in a recent
investigation exploring the impact of \textit{npe} URCA processes in BNS
simulations that the bulk-viscous approximation is valid in large parts
of the hypermassive neutron star (HMNS)~\cite{Most2022}. Furthermore,
results from moment-based treatments of neutrino
transport have not found evidence for large
out-of-thermodynamic equilibrium effects as those needed to produce an
effective bulk viscosity~\cite{Radice2022}, although being present in the
first few milliseconds after the merger~\cite{Espino2024b}.

In our recent work~\cite{Chabanov2023}, we obtain the first
quantitatively robust assessment of the impact of bulk viscosity on the
post-merger GW signal from BNS mergers by using the causal and
second-order formulation of dissipative hydrodynamics by M\"uller, Israel
and Stewart (MIS)~\cite{Mueller1967, Israel76, Israel1979, Hiscock1983,
  Rezzolla_book:2013, Gavassino2021,Most2021d,Camelio2022_a} (see
also~\cite{Camelio2022_b} for preliminary studies in spherical symmetry
and~\cite{Duez2004b, Radice2017, Shibata2017a, Shibata2017b, Duez2020}
for other interesting approaches to viscous dissipation in
general-relativistic simulations). Our microphysical description
consists of a bulk viscosity that is determined by direct and modified
Urca reactions that are responsible for achieving weak chemical
equilibrium in neutron-star matter. In addition, we vary the
composition of cold neutron-star matter above the nuclear-saturation
density in order to study systematically the impact of small
(realistic) and large (unrealistic) bulk viscosities. These scenarios
can be seen as corresponding to four different resonant maxima of the
so-called ``AC bulk viscosity'' \cite{Yang2023}
$\zeta_{_{\mathrm{AC}}}(\omega) := \zeta( 1 + \omega^2 \tau_{_{\Pi}}^2
)^{-1}$, where $\zeta$, $\tau_{_{\Pi}}$ and $\omega$ denote
respectively the bulk viscosity, the relaxation time and the angular
frequency of a periodic density oscillation \cite{Alford2018a,
  Sawyer1989}.

We find that large bulk viscosities are effective at damping the
collision-and-bounce oscillations of the stellar cores while, at the
same time, preserving the initial $m=2$ density deformations of the
binary.  The main consequences of the increase of the bar-mode
deformations constitute, on one hand, an increase of the efficiency of
energy and angular-momentum losses via GWs. On the other hand, the
stellar structure of the HMNS is modified and is characterised by a
more compact remnant with uniformly rotating core spinning faster than
in the inviscid case but having the same rotational kinetic energy. The
larger spinning frequency of the viscous remnants is reflected in a
larger value for the $f_2$ frequency in the post-merger power spectral
density (PSD). While the behaviour described above applies to all
viscous binaries, the differences between viscous and inviscid binaries
become significant only for the most extreme configurations.

This paper presents a comprehensive report on the numerical methods
applied in~\cite{Chabanov2023} where, however, we make use of a
particularly simplified microphysical model for the bulk
viscosity. More specifically, we here assume a \textit{constant} value
within the neutron stars which is set to zero below a certain threshold
density. Differently from our investigation in \cite{Chabanov2023},
where the transport coefficients arise consistently from microphysical
arguments leading to complex functions of density and temperature, this
is clearly an approximation. However, this treatment has the important
advantage of being independent of the EOS and thus more suitable for
serving as a numerical testbed.

We provide a detailed description of the algorithm solving the MIS
equations for bulk viscosity with particular focus on various limiting
procedures implemented in the inversion procedure from conservative to
primitive variables in Section~\ref{sec:methods}. Then, we provide two
test cases employing isolated neutron stars which are used to verify the
implementation. First, in Subsection~\ref{sec:num_vis} we simulate the
viscous damping of density perturbations of TOV solutions and second, in
Subsection~\ref{sec:mig} we analyze the migration of TOV solutions from
the unstable branch to the stable branch with bulk viscosity. The two
test cases provide a qualitative and quantitative assessment of the
implementation introduced before in Section~\ref{sec:methods}. Finally,
in Subsection~\ref{sec:binary} we present a detailed report on BNS
simulations using a constant bulk viscosity within the stars where we
report on thermal properties, density and temperature distributions as
well as dynamical mass ejection. Last but not least, the conclusion of
this work can be found in Section~\ref{sec:conc}.

Unless stated otherwise, we use geometrised units where the speed of
light $c=1$ and the gravitational constant $G=1$. Additionally, we set
the Boltzmann constant $k_B=1$ and express the temperature in units of
$\mathrm{MeV}$. Greek letters denote spacetime indices, \ie $\mu = 0,
1, 2, 3$, while Roman letters cover spatial indices only, \ie $i = 1,
2, 3$. Also, we make use of Einstein's summation convention and choose
the metric signature to be $(-,+,+,+)$. Bold symbols, such as
$\boldsymbol{g}$, refer to tensors of generic rank, while the symbol
$\boldsymbol{\nabla}$ denotes the covariant derivative with respect to
the four-metric $\boldsymbol{g}$.

\section{Formulation and methods}\label{sec:methods}

We solve the second-order MIS equations (see also~\cite{Disconzi2017,
Bemfica2019b, Kovtun2019, Hoult2020, Taghinavaz2020, Pandya2022b,
Bantilan2022} for recent progress on first-order theories) excluding the
effects of heat conduction and shear viscosity. To do so, we follow
closely the equations presented in~\cite{Chabanov2021} and refer to this
work for a more detailed presentation and discussion of the mathematical
formalism as well as the equations describing heat conduction and shear
viscosity. We also evolve the spacetime metric in order to explore
self-consistently the effects of bulk viscosity in BNS merger simulations.
In the following, we briefly summarize the evolved set of equations. 

We start by introducing the 3+1 split of the metric
\citep{Alcubierre:2008,Rezzolla_book:2013}
\begin{equation}
 d s^2=-(\alpha^2-\beta_{i}\beta^{i}) d t^2+2\beta_{i} d x^{i} d
t+\gamma_{ij} d x^{i} d x^{j}\,,
\end{equation}
where $\alpha$ is the so-called lapse function, the purely spatial vector
$\beta^{i}$ is the shift vector, and \( \gamma_{ij} \) denotes the
components of the purely spatial metric with
$\gamma:=\mathrm{det}(\gamma_{ij})$ defining its determinant. The fluid
is described by the baryonic rest-mass density current and the
energy-momentum tensor for a dissipative fluid with zero heat conduction
and shear viscosity, \ie only bulk viscosity, which in the Eckart frame
are given by
\begin{align}
J^{\mu} &= \rho u^{\mu}\,\\
T^{\mu \nu} &= e u^{\mu}u^{\nu} + (p+\Pi)h^{\mu \nu}\,,\label{eq:energy_momentum}
\end{align}
respectively. The thermodynamic quantities $\rho, e, p$ and $\Pi$ denote
the baryonic rest-mass density, the total energy density, the pressure in
thermodynamic equilibrium and the bulk-viscous pressure, respectively.
The bulk-viscous pressure can be understood as the correction to $p$
arising from out-of-equilibrium effects. The vector
$u^{\mu}=W(1/\alpha,v^{i}/\alpha-\beta^{i}/\alpha)^T$ denotes the fluid
four-velocity where $v^{i}$ are the components of the Eulerian
three-velocity together with the Lorentz factor
$W:=\sqrt{1-v_{i}v^{i}}^{\phantom{.}-1}$. At this point we also define
the \textit{``coordinate velocity''} \(V^{j}:=u^{j}/u^{t}=\alpha
v^{j}-\beta^{j}\). The conservation of baryon number as well as energy
and momentum read
\begin{align}
\nabla_{\mu}J^{\mu}&=0\,,\label{eq:mass_conserv}\\
\nabla_{\mu}T^{\mu \nu}&=0\,,\label{eq:en_conserv}
\end{align}
respectively. Together with the choice of an EOS, \ie $p=p(\rho,T)$,
where $T$ is the temperature of the fluid, we are left to specify a
constitutive relation for $\Pi$ in order to close the system of
equations. As in~\cite{Chabanov2021,Chabanov2023} we employ the
Maxwell-Cattaneo model
\begin{align}
\tau_{_{\Pi}} u^{\mu}\nabla_{\mu}\Pi &= \Pi_{\mathrm{NS}} -
\Pi\,,\label{eq:is_bulk}\\
\Pi_{\mathrm{NS}}&:=-\zeta \Theta\,,\label{eq:nsvalue}
\end{align}
where $\boldsymbol{u \cdot \nabla} = u^{\mu} \nabla_{\mu}$ refers to the
covariant derivative along the fluid four-velocity while
$\Pi_{\mathrm{NS}}$ and $\Theta :=\nabla_{\mu} u^{\mu}$ denote the
so-called Navier-Stokes (NS) value and the fluid expansion, respectively.
Further information on Eq.~\eqref{eq:is_bulk} can be found in
\cite{Chabanov2021}. It is interesting to note that recent work on the
relation between chemically reacting multi-fluids and bulk-viscous
single-fluids has shown that Eq.~\eqref{eq:is_bulk} can be used to
account even for the full out-of-equilibrium dynamics of a multi-fluid
with two species \cite{Gavassino2023,Yang2023}.

Altogether, we can write Eqs.~\eqref{eq:mass_conserv},
\eqref{eq:en_conserv} and \eqref{eq:is_bulk} in the flux-conservative
form 
\begin{equation}
\partial_{t}\boldsymbol{U}+
\partial_{i}\boldsymbol{F}^{i}(\boldsymbol{U})=\boldsymbol{S}(\boldsymbol{U},\partial_{\mu}\boldsymbol{U})\,,
\label{eq:pde_conservative}
\end{equation}
where $\boldsymbol{U}$, $\boldsymbol{F}^{i}$ and $\boldsymbol{S}$ denote
the state, flux and source vectors, respectively. For
Eqs.~\eqref{eq:mass_conserv}-\eqref{eq:is_bulk} their components are
given by
\begin{align}
&\boldsymbol{U} := \sqrt{\gamma}\begin{pmatrix}
	D \\[1em]
	S_{j} \\[1em]
	\tau \\[1em]
	D\Pi
	\end{pmatrix} =\sqrt{\gamma} \begin{pmatrix}
	\rho W \\[1em]
	\left( e+p+\Pi \right) W^2 v_{j} \\[1em]
	\left( e+p+\Pi \right) W^2 - \left( p + \Pi \right) - \rho W \\[1em]
	\rho W \Pi
	\end{pmatrix}\,,\label{eq:bulk_implement1}
\end{align}

\begin{align}
   \boldsymbol{F}^{i} :=\sqrt{\gamma} \begin{pmatrix}
    V^{i} D\\[1em]
	\alpha S^{i}_{\phantom{i}j} - \beta^{i}S_{j} \\[1em]
	\alpha (S^{i}-v^{i}D) - \beta^{i} \tau \\[1em]
	V^{i}D\Pi
	\end{pmatrix}\,, 
\end{align}

\begin{align}
    \boldsymbol{S} := \sqrt{\gamma} \begin{pmatrix}
	0 \\[1em]
	\frac{1}{2}\alpha S^{ik}\partial_{j} \gamma_{ik} +
S_{i}\partial_{j}\beta^{i} - (\tau+D) \partial_{j}\alpha \\[1em]
	\alpha S^{ik}K_{ij} - S^{j}\partial_{j}\alpha \\[1em]
	-(\alpha D/\tau_{_{\Pi}}W)
        \left[\zeta \left(
          \vartheta + \Lambda - KW \right)  + \Pi\right]
  \end{pmatrix}\,,\label{eq:bulk_implement2}
\end{align}
where we have introduced the following projections parallel and orthogonal to the
unit vector $n^{\mu}:=(1/\alpha,-\beta^{i}/\alpha)$: 
\begin{align}
D&:=-n_{\mu}J^{\mu}\,,\\
\tau&:=n_{\mu}n_{\nu}T^{\mu \nu}-D\,,\\
S_{j}&:=-\gamma_{j \mu}n_{\nu}T^{\mu \nu}\,,\\
S^{i}_{\phantom{i}j}&:=\gamma^{i}_{\phantom{i}\mu}\gamma_{j \nu}T^{\mu
\nu}\,.
\end{align}

Furthermore, it is useful to remind that the expansion scalar takes the
following form in the 3+1 split
\begin{align}
\Theta:&=\nabla_{\mu}u^{\mu} =\vartheta + \Lambda -
KW\,,\label{eq:3+1expansion}
\end{align}
with
\begin{align}
\vartheta &:=
D_{i}\left(Wv^{i}\right)=\frac{\partial_{i}
    \left(\sqrt{\gamma}Wv^{i}\right)}{\sqrt{\gamma}}\,,\\ \Lambda
&:= \frac{1}{\alpha}
  \left(\partial_{t}-\mathscr{L}_{\boldsymbol{\beta}}\right)
  W+Wv_{i}\hat{a}^{i}\,,\\
K &:= \gamma_{ij}K^{ij}\,.
\label{eq:Lambda}
\end{align}
The quantities $D_{i}$, $\mathscr{L}_{\boldsymbol{\beta}}$, $\hat{a}^{i}$ and
$K^{ij}$ denote the covariant derivative w.r.t. the purely spatial metric
$\gamma_{ij}$, the Lie derivative w.r.t.~the shift vector, the
acceleration of the normal observer
$\hat{a}^{i}:=\gamma^{ij}\partial_{j}\mathrm{ln}(\alpha)$ as well as the
so-called extrinsic curvature, respectively.

We close this section by introducing the definition of the relativistic
inverse Reynolds number based on \cite{Denicol2012b}
\begin{align}
\mathcal{R}^{-1}:=\frac{\Pi}{p+e}\,.\label{eq:reynolds}
\end{align}
From Eq.~\eqref{eq:reynolds} it becomes clear that the Reynolds number
measures the relative importance of inertial forces compared to viscous
or dissipative forces.

\subsection{Implementation}

The simulations reported below are obtained after solving Einstein's
equations together with the system of viscous second-order
general-relativistic hydrodynamics (GRHD) equations in
Eqs.~\eqref{eq:bulk_implement1}-\eqref{eq:bulk_implement2} via the
high-order high-resolution shock-capturing code
\texttt{FIL}~\cite{Most2019b, Most2019c}. The \texttt{FIL} code employs
fourth-order accurate finite-difference stencils in Cartesian coordinates
for the evolution of the constraint damping formulation of the Z4
formulation of the Einstein equations~\cite{Bernuzzi:2009ex,
Alic:2011a}. The equations of viscous GRHD are solved with a fourth-order
high-resolution shock-capturing scheme~\cite{DelZanna2007}, the
so-called ECHO scheme, built upon the open-source code
\texttt{IllinoisGRMHD}~\cite{Etienne2015}. More specifically, all flux
terms appearing in Eq.~\eqref{eq:pde_conservative} are discretized
following the ECHO scheme and the source terms appearing in the last
component of $\boldsymbol{S}$ in Eq.~\eqref{eq:bulk_implement2} are
discretized using finite differencing. Spatial derivatives are
discretized using fourth-order central finite differences while the temporal
derivative is calculated through first-order backward differencing
w.r.t.~the previous \textit{full} timestep. 

The discretization of the time derivative can be illustrated by the
following example: Let us assume we employ Heun's second-order method
with two stages and the solution should be advanced from $t$ to $t+\Delta
t$. The solution at $t-\Delta t$ is also available. Then, the first stage
employs the approximation 
\begin{align}
\left.\partial_t W\right|_{t} \approx \frac{W(t)-W(t-\Delta t)}{\Delta t}\,,
\end{align}
while in the second stage we use
\begin{align}
\left.\partial_t W\right|_{t+C \Delta t} \approx \frac{W(t+C \Delta
t)-W(t)}{C \Delta t}\,,
\end{align}
where $C=1$ for Heun's method. For higher-order multi-stage methods only
$C$ needs to be adjusted for each stage separately.
 
\subsection{Primitive inversion and limiting}

Typical methods for solving the equations of relativistic hydrodynamics
involve the conversion from the evolved conserved variables to the
primitive variables which has to be carried out numerically by using
root-finding algorithms. In this work, we augment the implementation of
the purely hydrodynamical algorithm from~\cite{Galeazzi2013} already
provided in \texttt{FIL} in order to include the bulk-viscous pressure
$\Pi$. As will be shown, almost all desirable properties, \ie existence
and uniqueness of solutions, can be
transferred to the viscous case if a suitable limiting is employed. In
the following, we just briefly review the most notable steps of the
primitive inversion algorithm presented in~\cite{Galeazzi2013}
and highlight the modifications which arise from the inclusion of bulk
viscosity. We omit all details related to the handling of the density
floor, the finite range of the EOS table as well as all rescalings that
follow from the bounds on the conserved and primitive variables. All of
these steps are not influenced by the inclusion of bulk viscosity except
that an atmosphere value for $\Pi$ needs to be specified which we choose
to be $\Pi_{\mathrm{atmo}}=0$.

We start by introducing the specific enthalpy $h:=(e+p)/\rho$, the
specific internal energy $\epsilon$, the norm of the Eulerian
three-velocity $v:=\sqrt{v_iv^i}$ and, as in~\cite{Galeazzi2013}, the
following definitions
\begin{align}
a' := \frac{p+\Pi}{\rho(1+\epsilon)}\,, \quad h' &:= h +
\frac{\Pi}{\rho}\,, \quad z := Wv\,,\\ q := \frac{\tau}{D}\,, \quad r &:=
\frac{S^{i}S_{i}}{D}\,, \quad k := \frac{r}{1+q}\,.
\end{align}
These definitions yield the following relations:
\begin{align}
z = \frac{r}{h'}\,, \quad \rho = \frac{D}{W}\,, \quad W = \sqrt{1+z^2}\,,
\end{align}
and most crucially
\begin{align}
\epsilon &= Wq - zr + W -1,\\
h' &= (1+\epsilon)(1+a')=(W-zk)(1+q)(1+a').\label{eq:hprime}
\end{align}
These expressions are exactly the same as in~\cite{Galeazzi2013}, if $a$
and $h$ are replaced by $a'$ and $h'$, respectively.

\subsubsection{Limiting\label{sec:limiting}}

We now impose the same requirements as formulated in~\cite{Galeazzi2013}
on the viscous fluid of our simulations, \ie the matter should satisfy
the dominant energy condition and the viscous sound speed has to be
smaller than the speed of light
\begin{align}
0\leq a' \leq 1, \label{eq:limit_aprime}\\
0\leq {c_s'}^2 < 1 \label{eq:limit_sound}.
\end{align}
The quantity ${c_s'}^2$ is the square of the sound speed including
modifications from bulk viscosity~\cite{Bemfica2019, Chabanov2021, Camelio2022_a}
\begin{align}
{c_s'}^2 = \frac{\zeta}{\tau_{_{\Pi}}}\frac{1}{\rho h'} + \left(\frac{\partial p}{\partial
  e}\right)_{\rho}+\frac{1}{h'} \left(\frac{\partial p}{\partial \rho
}\right)_{e}.
\end{align}
Because the bulk-viscous pressure $\Pi$ is treated as an independent quantity and
hence evolved along with the other typical hydrodynamic
variables, the fluid can easily evolve into states which violate the
conditions \eqref{eq:limit_aprime} and \eqref{eq:limit_sound}. Hence, we
formulate the following limiting strategy in order to avoid large
out-of-equilibrium contributions in the fluid:
\begin{enumerate}
\item[(i)] If $\Pi < 0$ and $\Pi < \sigma p$ with $-1 \leq \sigma \leq
0$, set $\Pi = \sigma p$. Note that for this range of the free parameter
$\sigma$ the condition \eqref{eq:limit_aprime} is automatically
fullfilled. We typically choose $\sigma=-0.9$ in order to avoid regions
where the fluid is pressureless, \ie $p+\Pi=0$.
\item[(ii)] If $\Pi > 0$ and $a' > 1$, set $\Pi = e-p$.
\item[(iii)] If ${c_s'}^2 > {c_{\mathrm{max}}}^2$, set 
\begin{align}\label{eq:causal_limit}
\tau_{_{\Pi}} = \frac{\zeta}{\rho h'}\left[{c_{\mathrm{max}}}^2-\left(\frac{\partial p}{\partial
  e}\right)_{\rho}-\frac{1}{h'} \left(\frac{\partial p}{\partial \rho
}\right)_{e}\right]^{-1}\,,
\end{align} 
where $c_{\mathrm{max}}$ is a free parameter between $0 \leq
{c_{\mathrm{max}}}^2 < 1$ in order explicitly ensure causality.
\end{enumerate}

\subsubsection{Bounds for the conserved variables}

Again following~\cite{Galeazzi2013}, we find the same lower bound on $q$
if the specific internal energy $\epsilon$ is required to be positive
\begin{align}
q = h'W-1-\frac{p+\Pi}{W\rho}\geq \epsilon \geq \epsilon_{\mathrm{min}}\,,
\end{align}
where $\epsilon_{\mathrm{min}}$ denotes the minimum specific internal
energy of the EOS. Additionally, we can also confirm that the
same relations and bounds for the total momentum density in terms of $k$
remain almost unchanged, \ie
\begin{align}
k(v,a') = v \frac{1+a'}{1+v^2a'}\,, \quad \frac{\partial}{\partial
a'}k(v,a') \geq 0\,,\\
0 \leq \frac{1}{2}k \leq v \leq k \leq \frac{2v}{1+v^2} < 1\,,
\label{eq:k_limit}
\end{align}
because we assume that the dominant energy condition is fullfilled.
Imposing an upper limit on the velocity $v$, which we denote as
$v_{\mathrm{max}}$, then automatically yields
\begin{align}
k < k_{\mathrm{max}} = \frac{2v_{\mathrm{max}}}{1+{v_{\mathrm{max}}}^2}\,.
\end{align}

\subsubsection{Root finding, existence and uniqueness}

We use the same root finding function as in~\cite{Galeazzi2013} except
that we substitute $h(z)$ by $h'(z)$, \ie 
\begin{align}
f(z)=z-\frac{r}{h'(z)}\,.
\end{align}
Note that we omit the $\sim$ superscript compared to~\cite{Galeazzi2013}
for simplicity. Furthermore, the limits are imposed at every step of the
root finding which means that when computing $a'(z)$ we impose the
requirements (i) and (ii). Then, the limited value for $a'(z)$ is used in
Eq.~\eqref{eq:hprime} to calculate $h'(z)$.

From Eq.~\eqref{eq:k_limit}
we find that the bracketing interval remains unchanged as well
\begin{align}
z_{-}=\frac{k/2}{\sqrt{1-k^2/4}}\,, \quad z_{+}=\frac{k}{\sqrt{1-k^2}}\,.
\end{align}
It is possible to show that $f(z)$ has one unique solution by following
the steps outlined in ~\cite{Galeazzi2013} and replacing $h$ and $a$ by
$h'$ and $a'$, respectively. There is only one notable difference: Due to
the fact the we limit $\Pi$ in the root finding process, it can only be
considered constant if the limiter is never applied. This leads to three
cases for which uniqueness can be established separately. Overall, we
obtain three different relations for the quantity $B$ defined through
\begin{align}
\frac{df}{dz}&=1-v^2B\,,\\
B&=a'\left[1+\frac{\partial \mathrm{ln}(1+a')}{\partial
\mathrm{ln}(1+\epsilon)}\right]+\frac{\partial \mathrm{ln}(1+a')}{\partial
\mathrm{ln}(\rho)}\,.
\end{align}
If $\Pi$ is assumed to be constant, then we obtain
\begin{align}
B_{_{\Pi}}=\left.\frac{\partial p}{\partial
e}\right|_{\rho}+\left.\frac{1}{h'}\frac{\partial p}{\partial
\rho}\right|_{e} \leq {c_s'}^2 < 1\,.
\end{align}
If $\Pi$ is limited from below, then we obtain
\begin{align}
B_{\mathrm{(i)}}=(1-\sigma)\left\{\left.\frac{\partial p}{\partial
e}\right|_{\rho}+\left.\frac{\rho}{(1-\sigma)p+e}\frac{\partial p}{\partial
\rho}\right|_{e}\right\} \leq {c_s}^2 < 1\,.
\end{align}
Finally, if $\Pi$ is limited from above, then we obtain
\begin{align}
B_{\mathrm{(ii)}}=a'=1\,.
\end{align}

\subsubsection{Reynolds number limits}

In this section we want to give an estimate for the minimum and maximum
inverse Reynolds numbers, $\mathcal{R}^{-1}_{\mathrm{min}}$ and
$\mathcal{R}^{-1}_{\mathrm{max}}$, respectively, \eg see
Eq.~\eqref{eq:reynolds}, achievable by our scheme. For that we employ the
simple ideal fluid EOS, \ie $p=(\Gamma_{\mathrm{id}}-1)\rho\epsilon$,
where $\Gamma_{\mathrm{id}}$
denotes the adiabatic index.  For an isentropic fluid which obeys the
ideal fluid EOS we can express the pressure additionally as
$p=\kappa \rho^{\Gamma_{\mathrm{id}}}$ such that the specific internal energy can be written
as $\epsilon=\kappa \rho^{\Gamma_{\mathrm{id}}-1}/(\Gamma_{\mathrm{id}}-1)$.  The minimum inverse Reynolds
number is given by (i):
\begin{align}
\mathcal{R}^{-1}_{\mathrm{min}}=\frac{\sigma
p}{e+p}=\sigma\frac{1}{\kappa^{-1}\rho^{1-\Gamma_{\mathrm{id}}}+1+(\Gamma_{\mathrm{id}}-1)^{-1}}.
\end{align}
A typical choice for test simulations of neutron stars is
$\Gamma_{\mathrm{id}}=2$,
$\kappa=100$ and $\rho=0.00128$ which yields
$\mathcal{R}^{-1}_{\mathrm{min}}\simeq -0.1$
for $\sigma=-0.9$.  The maximum inverse Reynolds number is given by (ii):
\begin{align}
\mathcal{R}^{-1}_{\mathrm{max}}=\frac{e-p}{e+p}=\frac{(\Gamma_{\mathrm{id}}-1)\rho^{1-\Gamma_{\mathrm{id}}}/\kappa-\Gamma_{\mathrm{id}}+2}{(\Gamma_{\mathrm{id}}-1)\rho^{1-\Gamma_{\mathrm{id}}}/\kappa+\Gamma_{\mathrm{id}}}.
\end{align}
Using the same choice for $\Gamma_{\mathrm{id}}$, $\kappa$ and $\rho$ we find
$\mathcal{R}^{-1}_{\mathrm{max}}\simeq 0.8$.

\subsection{Transport coefficients and transition to inviscid fluid}

As already mentioned, the scope of this is work is to examine, for the
first time, the impact of a causal and constant bulk viscosity
prescription on BNS merger simulations. However, as neutron star mergers
lead to violent shocks propagating outward through a decreasing density
profile, a constant value for the bulk viscosity which is applicable in
the high-density regime will lead to large inverse Reynolds numbers as
soon as the shock reaches sufficiently low densities. As a consequence,
our limiting procedure will have to be applied continuously in the low
density regime.  Thus, we implement a smooth transition zone between the
high-density bulk viscosity $\zeta_h$ and the low density bulk viscosity
$\zeta_l$ typically chosen to be $\zeta_l=0$. The high-density zone which
has $\zeta=\zeta_h$ occupies matter with densities $\rho\geq\rho_h$ while
the region with densities $\rho\leq \rho_l$ has $\zeta=\zeta_l$. The
functional behaviour of the bulk viscosity in the transition zone defined
as the density interval $\rho_l<\rho< \rho_h$ is expressed through a
cubic polynomial which ensures continuity of $\zeta$ and
$\mathrm{d}\zeta/\mathrm{d}\rho$ at $\rho_h$ and $\rho_l$. We have
found that using a cubic polynomial in place of a simple linear
interpolation leads to bulk viscosity with better functional
behaviours.

Overall, our prescription can be summarized as follows:
\begin{align}
\zeta(\rho)=\left\{\begin{array}{ccc}
\zeta_l & \mathrm{if} & \rho \leq \rho_l\,,\\
a\rho^3+b\rho^2+c\rho+d & \mathrm{if} & \rho_l < \rho < \rho_h\,, \\
\zeta_h & \mathrm{if} & \rho \geq \rho_h\,,\label{eq:zeta_bound}
\end{array}\right.
\end{align}
where the coefficients of the cubic polynomial are given by
\begin{align}
a&=2(\zeta_l-\zeta_h)/(\rho_h-\rho_l)^3\,,\\
b&=3(\rho_l+\rho_h)(\zeta_h-\zeta_l)/(\rho_h-\rho_l)^3\,,\\
c&=6\rho_h\rho_l(\zeta_l-\zeta_h)/(\rho_h-\rho_l)^3\,,\\
d&=\left[(3\rho_h\rho_l^2-\rho_l^3)\zeta_h+(\rho_h^3-3\rho_h^2\rho_l)\zeta_l\right]/(\rho_h-\rho_l)^3\,.
\end{align}
Additionally, the relaxation time is set to be constant using a
prescription with a linear interpolation in the transition zone
\begin{align}
\tau_{_{\Pi}}(\rho)=\left\{\begin{array}{ccc}
\tau_l & \mathrm{if} & \rho \leq \rho_l\,,\\
\tau_h-m(\rho_h-\rho) & \mathrm{if} & \rho_l < \rho < \rho_h\,, \\
\tau_h & \mathrm{if} & \rho \geq \rho_h\,,
\end{array}\right.
\end{align}
where 
\begin{align}
m=(\tau_h-\tau_l)/(\rho_h-\rho_l)\,.
\end{align}
In contrast to the bulk viscosity $\zeta$, we found that a linear
functional form for $\tau$ in the transition zone is sufficient to
ensure numerical stability. We typically choose $\tau_h$ to be $\sim
1.1\Delta t_{\mathrm{min}}$ and $\tau_l$ to be $\sim 1.1\Delta
t_{\mathrm{max}}$, where $\Delta t_{\mathrm{min}}$ and $\Delta
t_{\mathrm{max}}$ are the minimum and maximum timesteps in the
simulation, respectively. If the neutron stars are fully covered by the
highest refinement level and the hydrodynamical timescale is
well-resolved by $\Delta t_{\mathrm{min}}$, this prescription for the
relaxation time ensures that $\Pi\simeq \Pi\ns = -\zeta \Theta$ and the
avoidance of stiff sources. Note that at the same time causality is
guaranteed through Eq.~\eqref{eq:causal_limit} which can lead to a local
increase of $\tau_{_\Pi}$.

\begin{figure}
\includegraphics[width=0.49\textwidth]{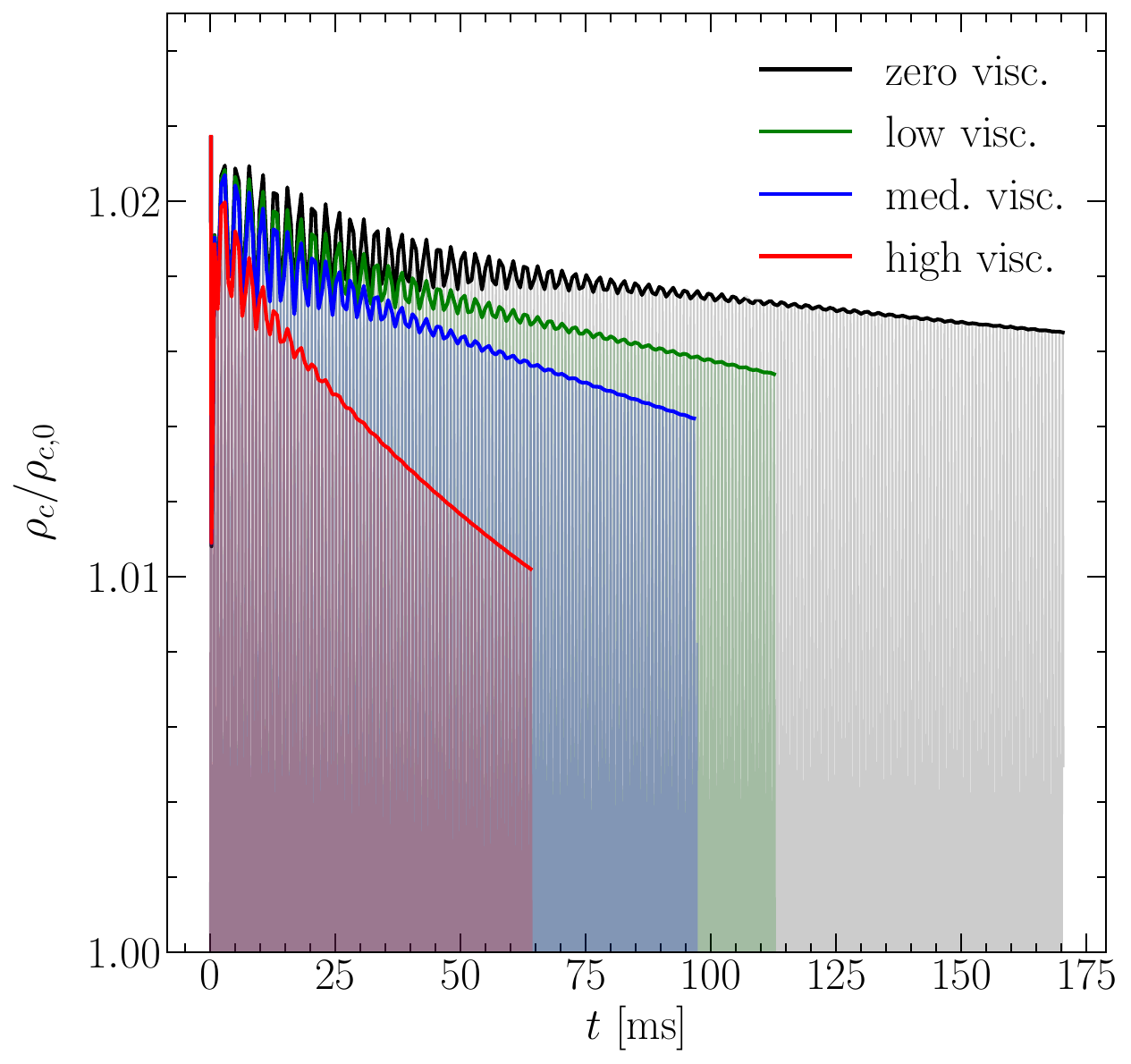}
\caption{Shown in solid transparent lines is the timeseries of the
central rest-mass density normalized by its value at $t=0$
for the zero (black line), low (green line),
medium (blue line) and high (red line) viscosity case. We present only
results from the highest resolution simulations, \ie $\Delta x \approx
207~\mathrm{m}$. Solid thick lines of the corresponding color show
the envelope of $\rho_{c}/\rho_{c,0}$ which illustrates the impact of
different bulk viscosities on the damping time of linear density
oscillations.  
\label{fig:osc_nofit}
}
\end{figure}

\begin{figure}
\includegraphics[width=0.49\textwidth]{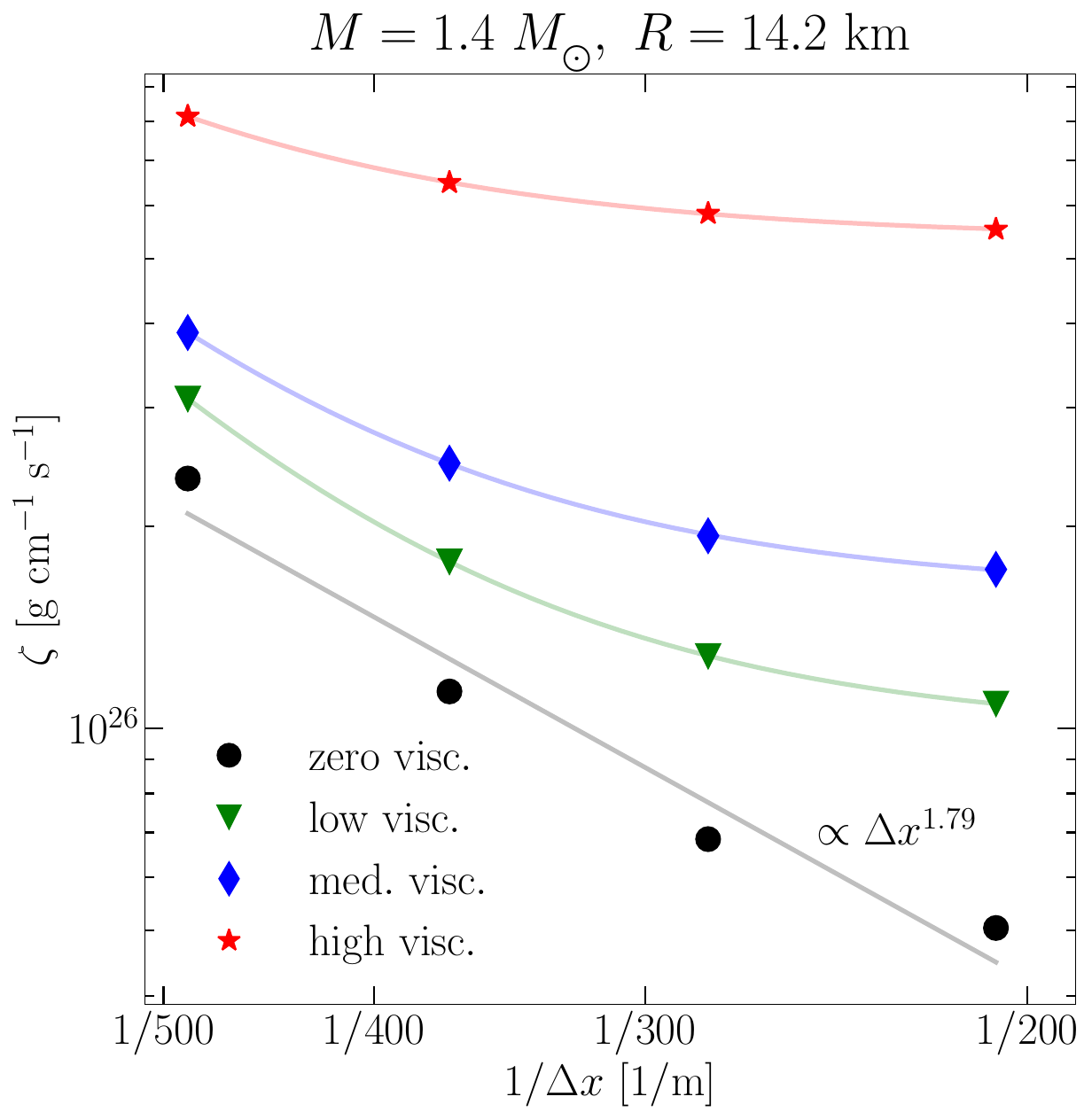}
\caption{Different symbols denote measured values of the bulk viscosity
  calculated by employing the damping time of central density
  oscillations of perturbed TOV solutions. We present the zero, low,
  medium and high-viscosity case using black circles, green triangles,
  blue diamonds and red stars, respectively. Transparent solid lines of
  the corresponding color display fits to the measured values as a
  function of grid resolution. The transparent black line shows a linear
  fit of $\log[\zeta]$ as a function of $\log[\Delta x]$ for the
  zero-viscosity case. The parameters of the TOV solution are given in
  the title.
\label{fig:num_vis}
}
\end{figure}

\section{Results}
\subsection{Oscillating neutron stars and numerical viscosity}
\label{sec:num_vis}

We now move on to the first numerical experiments involving isolated
neutron stars in order to test the implementation. Specifically, we
investigate the damping of radial perturbations of a neutron star in full
general relativity.  In this test, we set out to measure the numerical
and physical viscosity of our code. The procedure has been laid out
in~\cite{CerdaDuran2010} and we follow it for the most part except of
some minor differences. A detailed account of the measurement procedure
employed in this work is presented in
Appendix~\ref{sec:measurement_proc}. In short, we extract the damping
time of central density oscillations from our simulations and use it to
obtain an estimate for $\zeta_h$.

We use simulations of isolated TOV stars using a simple hybrid polytropic
EOS~\cite{Rezzolla_book:2013}, \ie
\begin{align}
p=\kappa\rho^{\Gamma}+(\Gamma_{\mathrm{th}}-1)\rho\epsilon_{\mathrm{th}}\,,
\label{eq:polyeos}
\end{align}
with $\kappa=100,~\Gamma=2,~\Gamma_{\mathrm{th}}=1.1$ and a central
density of $\rho_c=1.28\times10^{-3}~M_{\odot}^{-2}\approx 7.91\times
10^{14}~\mathrm{g}~\mathrm{cm}^{-3}$. The quantity $\epsilon_{\mathrm{th}}$
denotes the thermal component of the specific internal energy. This setup
yields a $M=1.4~M_{\odot}$ star, where $M$ denotes the
Arnowitt-Deser-Misner (ADM) mass, with a radius of $R=14.2~\mathrm{km}$.
This configuration is well-explored with known pulsation
frequencies~\cite{Font02c}. We use such a low $\Gamma_{\mathrm{th}}$
because we have found that the drift of $\rho_{c}$
can be reduced, if $\Gamma_{\mathrm{th}}$ assumes values close to but
larger than 1. 

Overall, we carry out 16 simulations which correspond to four different
bulk viscosities and four different resolutions. We vary the bulk
viscosity between $\zeta_h \in [0, \sim 9.42 \times 10^{25}, \sim 1.98
\times 10^{26}, \sim 8.20 \times 10^{26}]~\mathrm{g}~
\mathrm{s}^{-1}~\mathrm{cm}^{-1}$, which are denoted as zero, low, medium
and high-viscosity cases, respectively. The resolution on the finest
refinement level varies between $\Delta x \in \{\sim 207, \sim 281, \sim
369, \sim 487\}~\mathrm{m}$. Furthermore, we set $\zeta_l=0$,
$\rho_h=1.28\times10^{-4}~M_{\odot}^{-2}\approx7.91\times
10^{13}~\mathrm{g}~\mathrm{cm}^{-3}$ and
$\rho_l=1.28\times10^{-7}~M_{\odot}^{-2}\approx 7.91\times
10^{10}~\mathrm{g}~\mathrm{cm}^{-3}$. 

We initialize the stars by using a perturbation in the form of the
fundamental radial eigenmode. The perturbation produces oscillations in
the central density on the order of $\sim 2~\%$, \eg see
Fig.~\ref{fig:osc_nofit} which shows damped density oscillations of
$\rho_{c}$ normalized by its initial value at $t=0$, \ie
$\rho_c/\rho_{c,0}$ with $\rho_{c,0}:=\rho_c(t=0)$. We present the zero,
low, medium and high-viscosity case using a black, green, blue and red
line, respectively. All cases shown in Fig.~\ref{fig:osc_nofit} are
simulated at the highest resolution, \ie $\Delta x \approx
207~\mathrm{m}$. Such a small oscillation amplitude ensures that our
simulations remain in the linear regime of the coupled viscous
hydrodynamic and gravitational equations. The perturbation is initialized
by using the \texttt{PizzaTOV} thorn which is part of the publicly
available thorns accompanying the \texttt{WhiskyTHC}
code~\cite{Radice2012a,Radice2013b} built upon the \texttt{Einstein
  Toolkit}~\cite{Loffler:2011ay}.

During the simulations the perturbed stars oscillate almost entirely in
their fundamental radial eigenmode whose amplitude decays over time.
However, especially for the zero-viscosity case, the oscillations are
weakly contaminated by higher-order modes at early times, as can be seen
in Fig.~\ref{fig:osc_nofit}. The damping time of the amplitude is related
to the magnitude of $\zeta_h$ such that larger bulk viscosities lead to
shorter damping times.

The amplitude of the oscillations is subject to two damping mechanisms,
namely numerical and physical damping. The numerical damping stems from a
finite grid resolution and gives rise to a so-called numerical viscosity
which vanishes in the limit of infinite grid resolution. The physical
damping is related to our implementation of bulk viscosity and is
independent of grid resolution. Figure~\ref{fig:num_vis} shows the values
of $\zeta$ obtained through our measurement procedure for all 16
simulations denoted by different symbols. Additionally, we present fits
to the measured viscosities as a function of grid resolution in
transparent lines. For nonzero input viscosities, \ie nonzero $\zeta_h$,
we use the formula 
\begin{align}
\zeta(\Delta x)=\zeta_a+\zeta_s(\Delta x/\lambda)^{p_h}\,,\label{eq:numvisfit}
\end{align}
where $\zeta_a$ is the asymptotically measured viscosity at infinite grid
resolution, $\zeta_s$ is a numerical viscosity scale, $p_h$ is the
convergence order and $\lambda=2\pi c_s \omega^{-1}$ is the wavelength of
the perturbation with $\omega$ being its angular frequency. For the zero
viscosity case we employ a linear fit of $\log[\zeta]$ as a function of
$\log[\Delta x]$ which is equivalent to setting $\zeta_a=0$ in
Eq.~\eqref{eq:numvisfit}. We make this choice because keeping
$\zeta_a$ as an independent fitting coefficient in the zero-viscosity
case, which could possibly assume values not equal to zero, would lead
to overfitting of the data and misleading results for $p_h$. The results
for all fitting coefficients are shown in Table~\ref{tab:fits}.

\begin{table}
\centering
\begin{tabular}{lcccc}
\hline
\hline
Model
& $\zeta_a$ & $\zeta_s$ & $p_h$ \\
\hline
& $[\mathrm{g}~\mathrm{s}^{-1}~\mathrm{cm}^{-1}]$ & 
$[\mathrm{g}~\mathrm{s}^{-1}~\mathrm{cm}^{-1}]$ & 
 \\
\hline
zero visc. & $--$ & $4.4 \times 10^{29}$ & $1.79$ \\[3pt] 
low visc. & $9.89\times 10^{25}$ & $8.87 \times 10^{32}$ & $3.57$ \\[3pt]
med. visc. & $1.6\times 10^{26}$ & $5.65 \times 10^{32}$ & $3.45$ \\[3pt]
high visc. & $5.36\times 10^{26}$ & $2.75 \times 10^{32}$ & $3.23$ \\
\hline
\hline
\end{tabular}
\caption{Fitting results to the measured bulk-viscosity values
    presented in Fig.~\ref{fig:num_vis} using Eq.~\eqref{eq:numvisfit}
    with $\lambda \approx 3.48 \times 10^{4}~\mathrm{m}$.}
\label{tab:fits} 
\end{table}

\begin{figure*}
\includegraphics[width=0.315\textwidth]{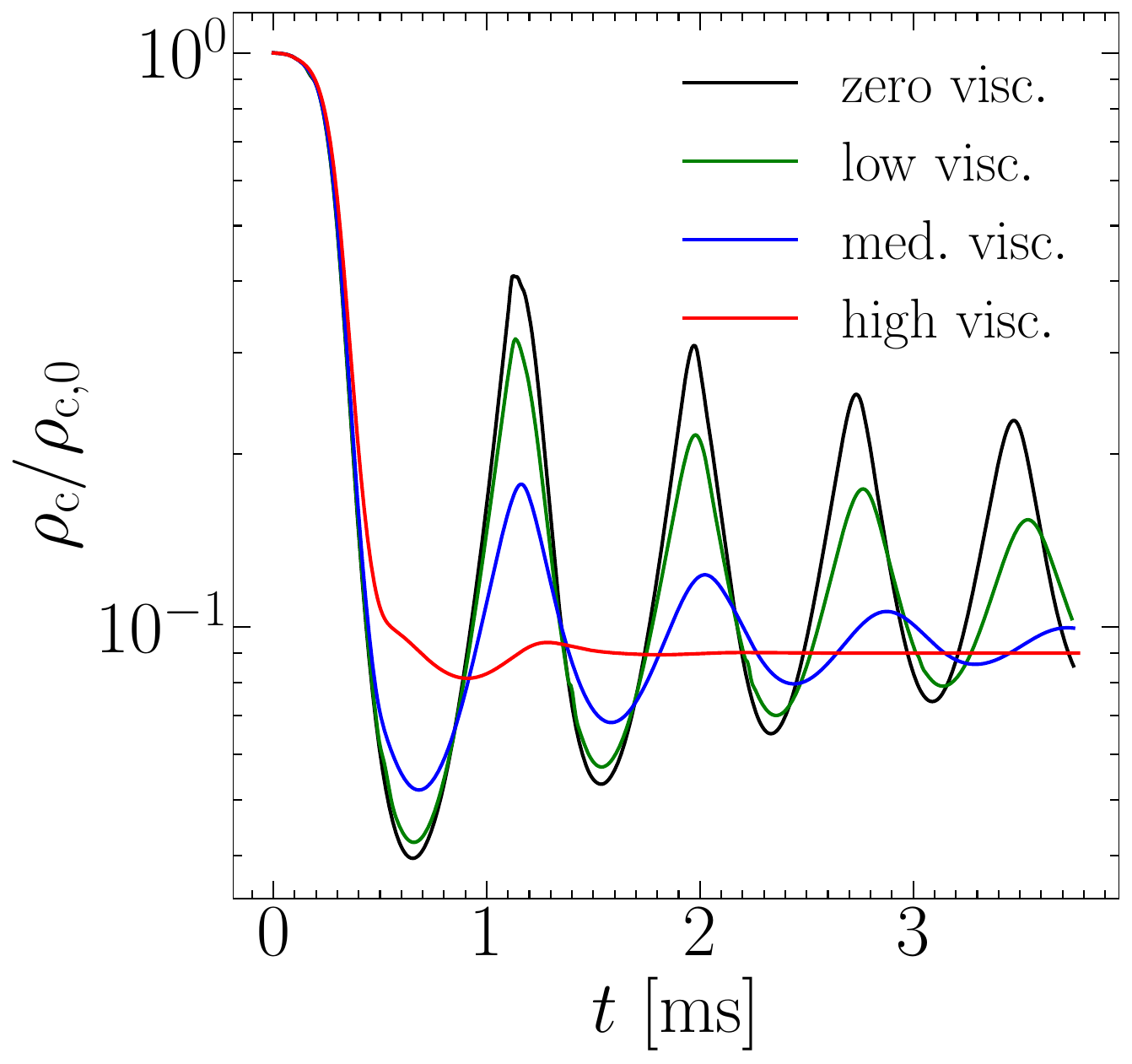}
\hskip 0.3cm
\includegraphics[width=0.315\textwidth]{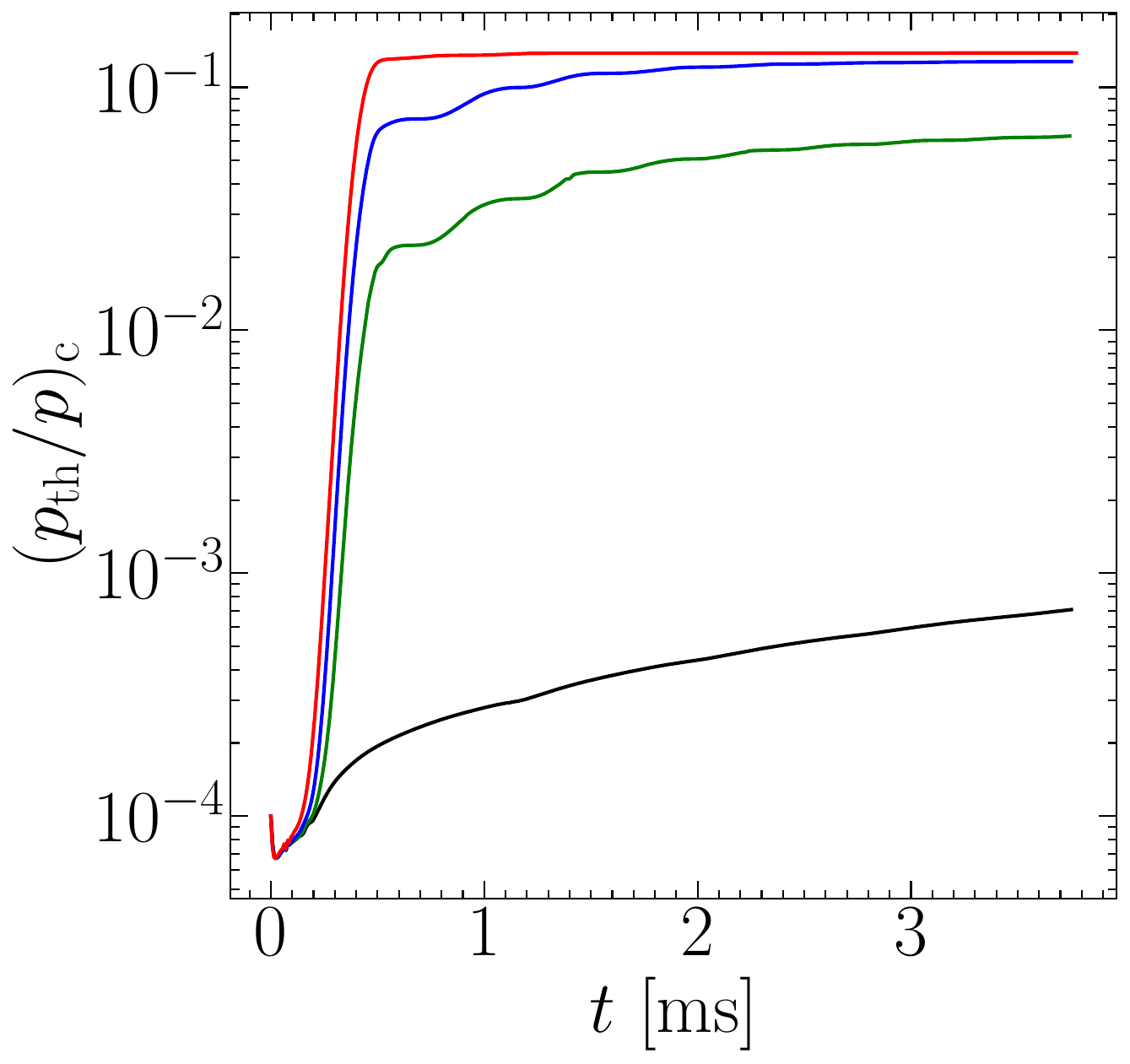}
\hskip 0.3cm
\includegraphics[width=0.315\textwidth]{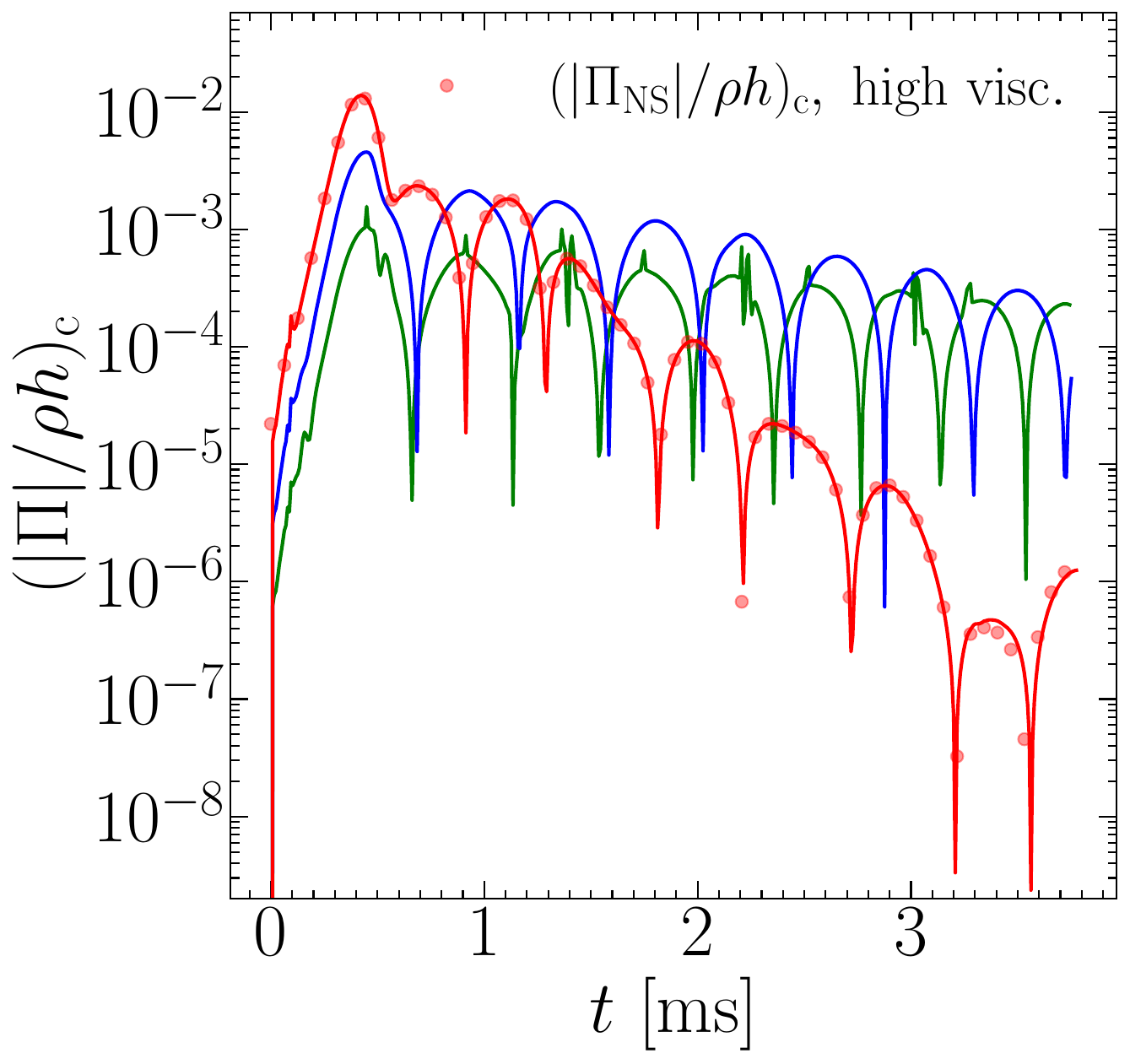}
\caption{Shown are the normalized density (left panel), the ratio of the
thermal pressure over the total pressure (middle panel) and the absolute
inverse Reynolds number (right panel), \ie the ratio of the absolute
value of the bulk-viscous pressure over the enthalpy density, in the
center of the migrating neutron star for the zero (black lines), low
(green lines), medium (blue lines) and high (red lines) viscosity case. The
transparent red symbols in the right panel denote the ratio of the
absolute NS value over the enthalpy density for the high-viscosity case.
\label{fig:mig_central}
}
\end{figure*}

There is a number of observations that can be made. First, all
nonzero-viscosity cases lead to measured bulk viscosities that decrease
as the resolution is increased (the cell size is smaller) and that
asymptote to a constant value. This confirms that, for high enough
resolutions, the damping of the corresponding density oscillations is
independent of the employed cell size.

\begin{figure*}
\includegraphics[width=0.32\textwidth]{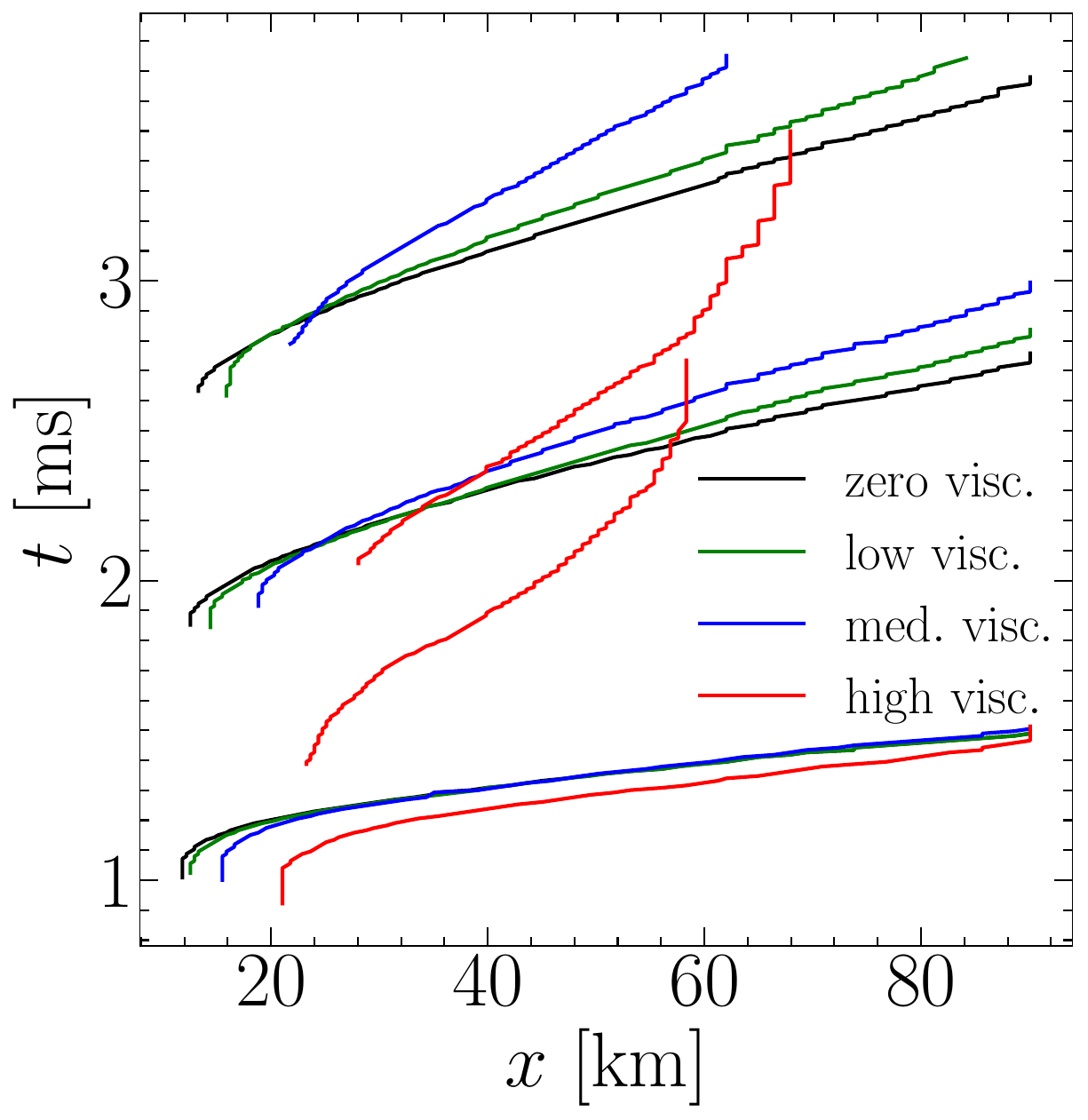}
\hskip 0.3cm
\includegraphics[width=0.65\textwidth]{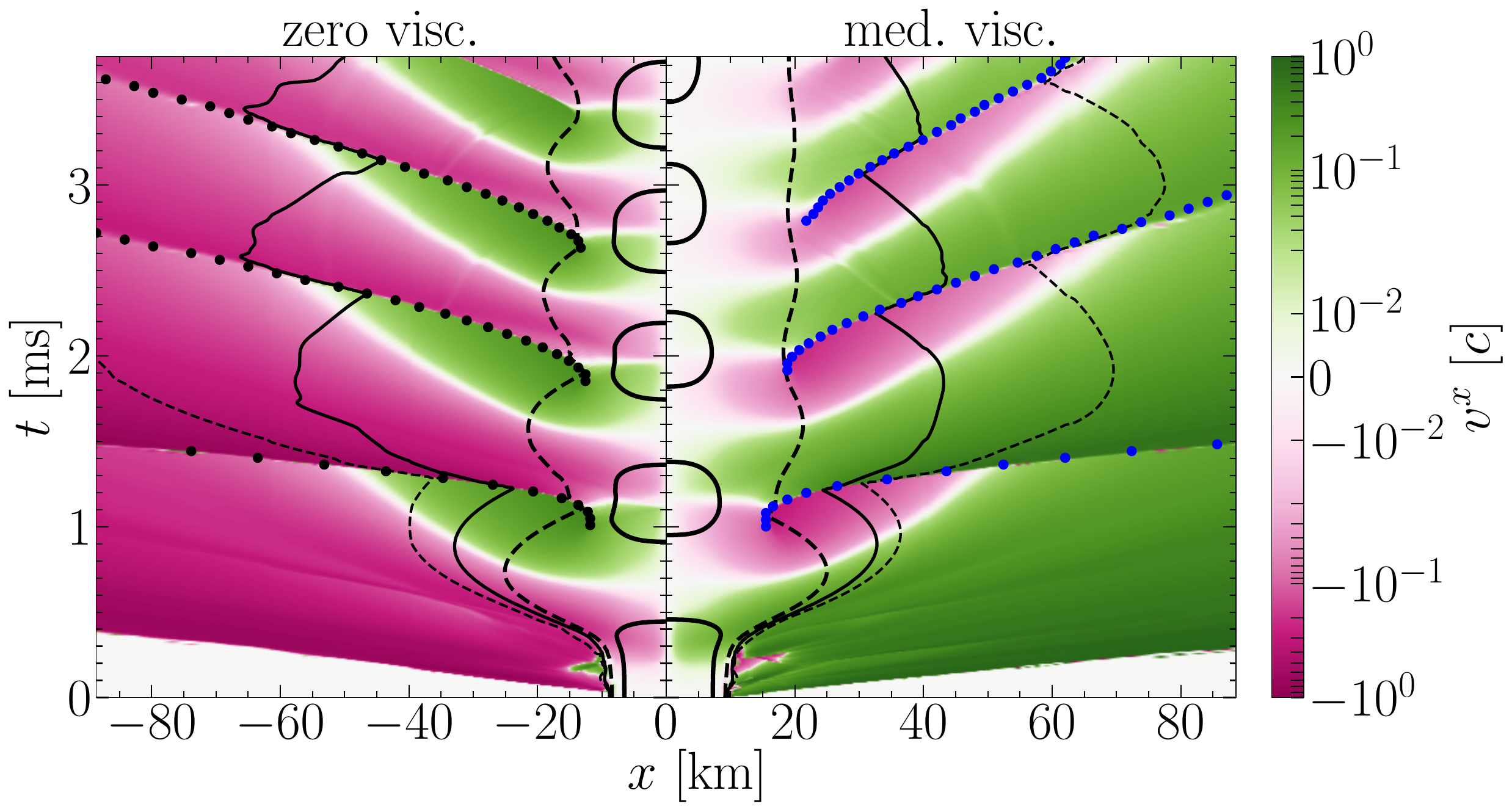}
\caption{\textit{Left}: Spacetime diagram of the evolution of the shock
  fronts on the $x$-axis which develop at the neutron star surface as a
  result of the violent nonlinear oscillations during the migration of
  the neutron star. We present the zero, low, medium and high viscosity
  case using black, green, blue and red lines, respectively.
  \textit{Right}: Spacetime diagrams of the evolution of the
  $x$-component of the Eulerian three-velocity on the $x$-axis for the
  zero (left half) and medium-viscosity (right half) case. The symbols
  denote the evolution of the corresponding shock fronts already
  presented in the \textit{left} panel. The inner solid, inner dashed,
  outer solid and outer dashed lines denote density contours at $\rho =
  4.5 \times [10^{14},10^{12},10^{10},10^{8}]\, \mathrm{g} \,
  \mathrm{cm}^{-3}$, respectively. Note that the inner solid line
  represents $\rho_h$ while the outer solid line corresponds to $\rho_l$,
  see also Eq.~\eqref{eq:zeta_bound}.
\label{fig:mig_shocks}
}
\end{figure*}

Bearing in mind that the measured viscosity $\zeta_a$ and the input
viscosity $\zeta_h$ are distinct but mathematically and physically
related, when comparing the values of $\zeta_a$ with those of $\zeta_h$
we find a good agreement notwithstanding the approximations involved in
deriving the formulas for the measurement of $\zeta_a$, see \eg
Appendix~\ref{sec:measurement_proc}. More specifically, we find an
agreement of $\lesssim 6~\%$, $\lesssim 20~\%$ and $\lesssim 35~\%$ for
the low, medium and high-viscosity case, respectively. The reasons for
the observed differences can be numerous, \eg the usage of Newtonian
formulas, a varying numerical and physical viscosity profile within the
star, or the influence of discretization errors from the neutron star
surface. An additional, and possibly strong, assumption is that the
Newtonian damping time calculated through the relation between
oscillatory kinetic energy and the corresponding energy dissipation in
Eq.~\eqref{eq:damping_tau} of Appendix~\ref{sec:measurement_proc}
continues to give a good approximation for the damping time of the
density oscillations in fully general-relativistic evolutions of TOV
stars. In summary, given the perturbative nature of our study, we have
found that Eq.~\eqref{eq:damping_tau} approximates reasonably well the
damping of $\rho_{c}$.

Second, we observe slow convergence of the numerical viscosity in the
zero-viscosity case. The \texttt{FIL} code employs 4th-order accurate
methods for the discretization in space, see~\cite{Most2019b}, while the
employed time integrator is of order three. This means that the observed
convergence order of $\sim 1.79$ is far below the formal one. Note that
the discretization of $\Theta$ in Eq.~\eqref{eq:is_bulk} does not impact
the convergence order for the zero-viscosity case. We suspect that this
behaviour is related to the influence of discretization errors from the
neutron star surface whose ill-balanced density and pressure gradients
manifest as effective discontinuities in the numerical solution. This
leads to a reduction of the observed convergence order. However, it is
interesting to note that we were able to recover the correct formal
convergence order in similar simulations employing smaller neutron stars,
\ie TOV solutions with a smaller radius. We report on these findings in
Appendix~\ref{sec:small_star}.  

\subsection{Migration of neutron stars and propagation of shock waves}
\label{sec:mig}

In this test, we simulate a migrating TOV star~\cite{Font02c} with four
different values for the bulk viscosity, \ie $\zeta \in [0, 4 \times
  10^{28}, 2 \times 10^{29}, 10^{30}]\, \mathrm{g} \, \mathrm{s}^{-1} \,
\mathrm{cm}^{-1}$, which are denoted as zero, low, medium and high
viscosity cases, respectively. We use the same hybrid polytropic EOS as
in the numerical viscosity measurement, see Eq.~\eqref{eq:polyeos}, and
choose $\kappa=100$ and $\Gamma=\Gamma_{\mathrm{th}}=2$. As
in~\cite{Font02c} the stars are initialized to have a central rest-mass
density of $\rho_c=8 \times 10^{-3}~M_{\odot}^{-2}\approx 4.94 \times
10^{15}~\mathrm{g}~\mathrm{cm}^{-3}$ which results in a mass of
$M=1.447\, M_{\odot}$. Furthermore, the computational grid has outer
boundaries at $64\,M_{\odot}\simeq \,{\rm 95\, km}$ in the three spatial
directions and we employ a $z$-symmetry in the equatorial plane. We use
four refinement levels with a factor of two refinement; the last level,
which has a width of $24~M_{\odot}\simeq \,{\rm 35\, km}$ has a grid
spacing of $\Delta x = 0.25\,M_{\odot}\simeq 370\,\mathrm{m}$.

We start by describing the evolution of the characteristic central
quantities shown in Fig.~\ref{fig:mig_central}. The left panel shows the
evolution of the central rest-mass density normalized by its initial
value at $t=0$ for all four cases in different colors. First, as
expected, the magnitude of $\zeta$ has a strong influence on the damping
time of the nonlinear central density oscillations. As already observed
in the numerical viscosity measurement, higher values for the bulk
viscosity lead to smaller values for the damping time such that the
nonlinear oscillations of $\rho_c$ decay more efficiently.

Additionally, we observe that the average value around which these
oscillations occur decreases with increasing bulk viscosity. This can be
understood by looking at the middle panel in Fig.~\ref{fig:mig_central}
which shows the evolution of the central thermal pressure component
$p_{\mathrm{th}}:=(\Gamma_{\mathrm{th}}-1)\rho \epsilon_{\mathrm{th}}$
normalized by the central total pressure $p$. Higher values for $\zeta$
lead to a more rapid initial increase and higher asymptotic values of
$p_{\mathrm{th}}/p$ in the center of the star. As a result, the migrating
TOVs with a larger value for the bulk viscosity have larger thermal
support which leads to less compact asymptotic states with a smaller
central rest-mass density.

Finally, the right panel of Fig.~\ref{fig:mig_central} shows the absolute
value of the relativistic inverse Reynolds number in the center of the
stars. The quantity $\mathcal{R}^{-1}$ can be used to gain intuition into
estimating the importance of viscous effects in a generic relativistic
hydrodynamic flow while also quantifying the applicability of the system
of equations employed to describe out-of-equilibrium effects.  Regarding
the former, we observe a dramatic influence of viscous effects on the
evolution of the central-rest mass density for a flow with
$|\mathcal{R}^{-1}| \gtrsim 10^{-2}$. Regarding the latter, we conclude
that our simulations are still in a regime where the equations employed
remain in their domain of applicability due to the low inverse Reynolds
number. Additionally, the red transparent circles show the inverse
Reynolds number computed from the NS value for the high-viscosity case.
We observe good agreement between the true bulk-viscous pressure $\Pi$
and $\Pi\ns$ which means that our simulations deviate only very weakly
from a first-order formulation of relativistic dissipative hydrodynamics
while at the same time being fully causal. Note that NS values are
presented only for the high-viscosity case to avoid overcrowding the
figure. However, we observe the same agreement for the low and
medium-viscosity case.

Now, we move on to the evolution of shock waves commonly observed in
migration tests. Our results are visualized in
Fig.~\ref{fig:mig_shocks}.  The left panel of Fig.~\ref{fig:mig_shocks}
shows a spacetime diagram of the evolution of the first three shock
fronts appearing for each value of the bulk viscosity $\zeta$ on the
x-axis of our simulations. The right panel shows spacetime diagrams of
the Eulerian three-velocity on the x-axis, \ie $v^x(x, y=0, z=0)$, for
the zero (left half) and medium viscosity (right half)
case. Additionally, we use black contour lines to plot the evolution of
the density contours at $\rho \in 4.5 \times
[10^{14},10^{12},10^{10},10^{8}]\, \mathrm{g} \, \mathrm{cm}^{-3}$
using solid, dashed, solid and again dashed line styles,
respectively. Note that the inner solid line represents $\rho_h$ while
the outer solid line corresponds to $\rho_l$, see
Eq.~\eqref{eq:zeta_bound}. The colored markers denote the locations of
the first three shock fronts for each of the two simulations,
respectively.


First, we observe that a larger value for the bulk viscosity leads to a
shock front which develops at larger distances from the core. In order to
understand this, we need to describe briefly the generic mechanism of
shock-front formation. Prior to the development of the shock front, the
neutron star is contracting and matter at larger distances is radially
infalling. As soon as the contraction of the neutron star is decelerated
as the result of increasing pressure gradients, radially infalling
material hits the neutron star surface and gets shocked.  Consequently,
an outward propagating shock is generated as soon as the shock is strong
enough to overcome the ram pressure generated by the radially infalling
material. 

\begin{figure*}
\includegraphics[width=0.315\textwidth]{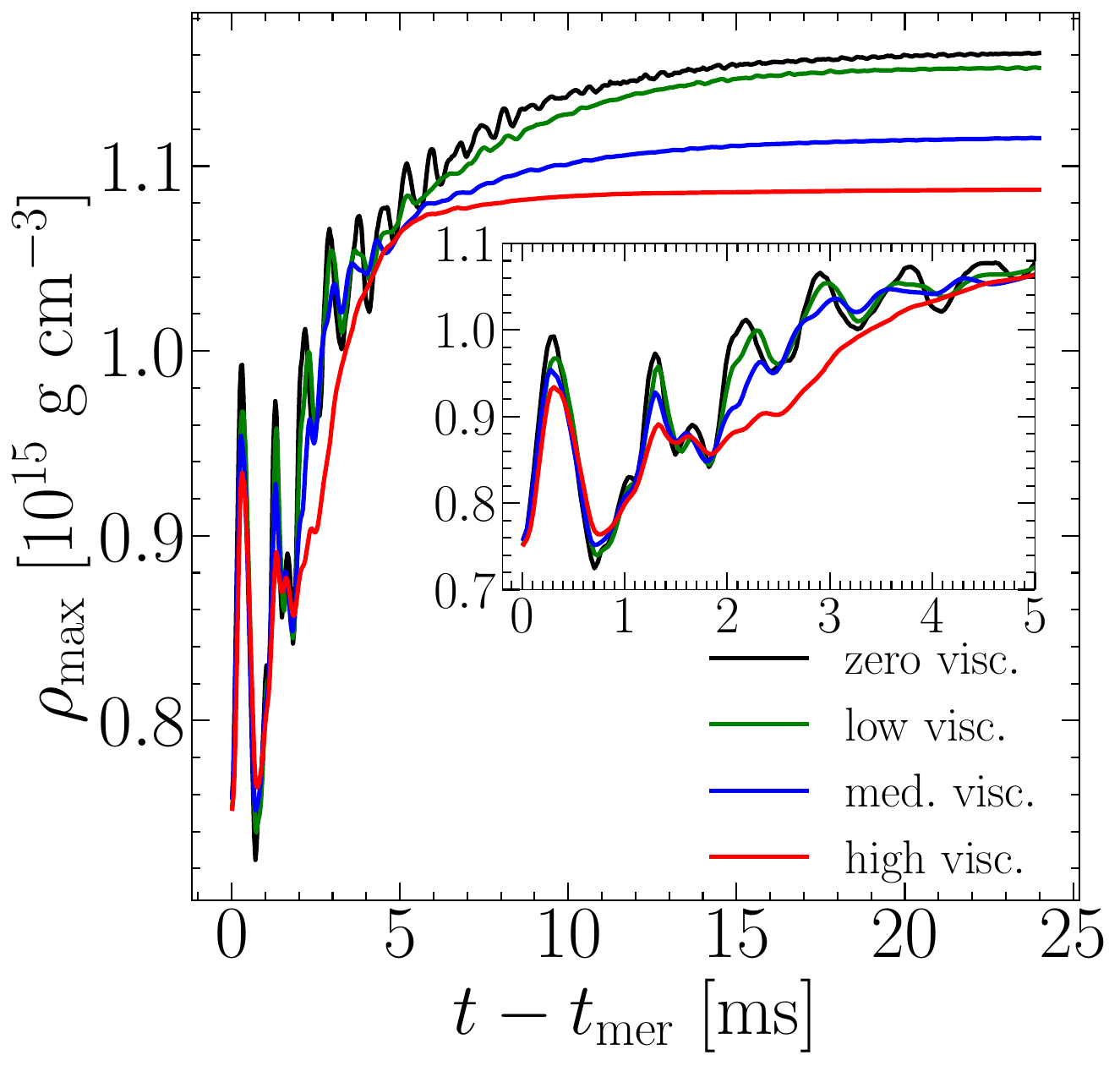}
\hskip 0.3cm
\includegraphics[width=0.315\textwidth]{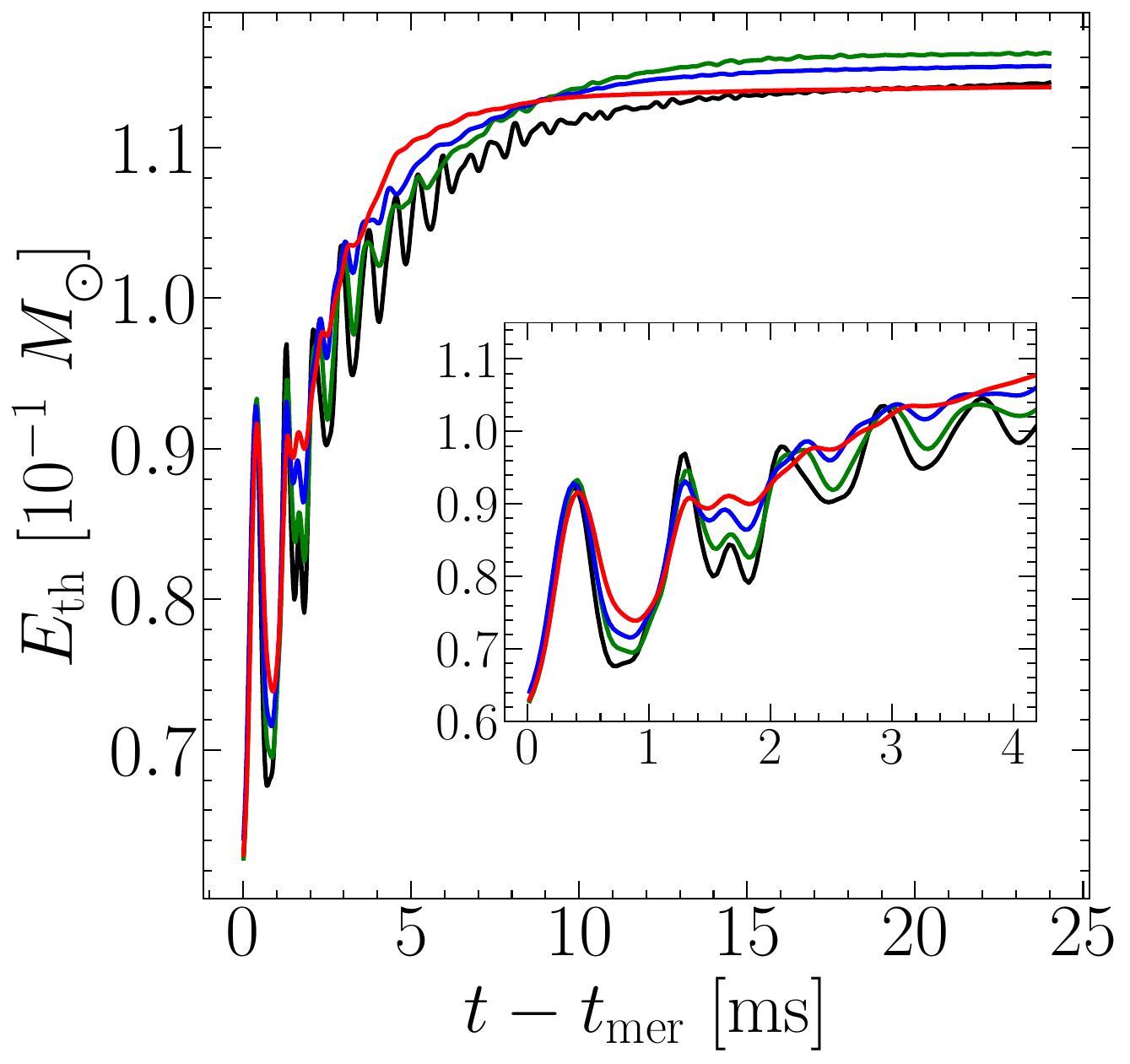}
\hskip 0.3cm
\includegraphics[width=0.315\textwidth]{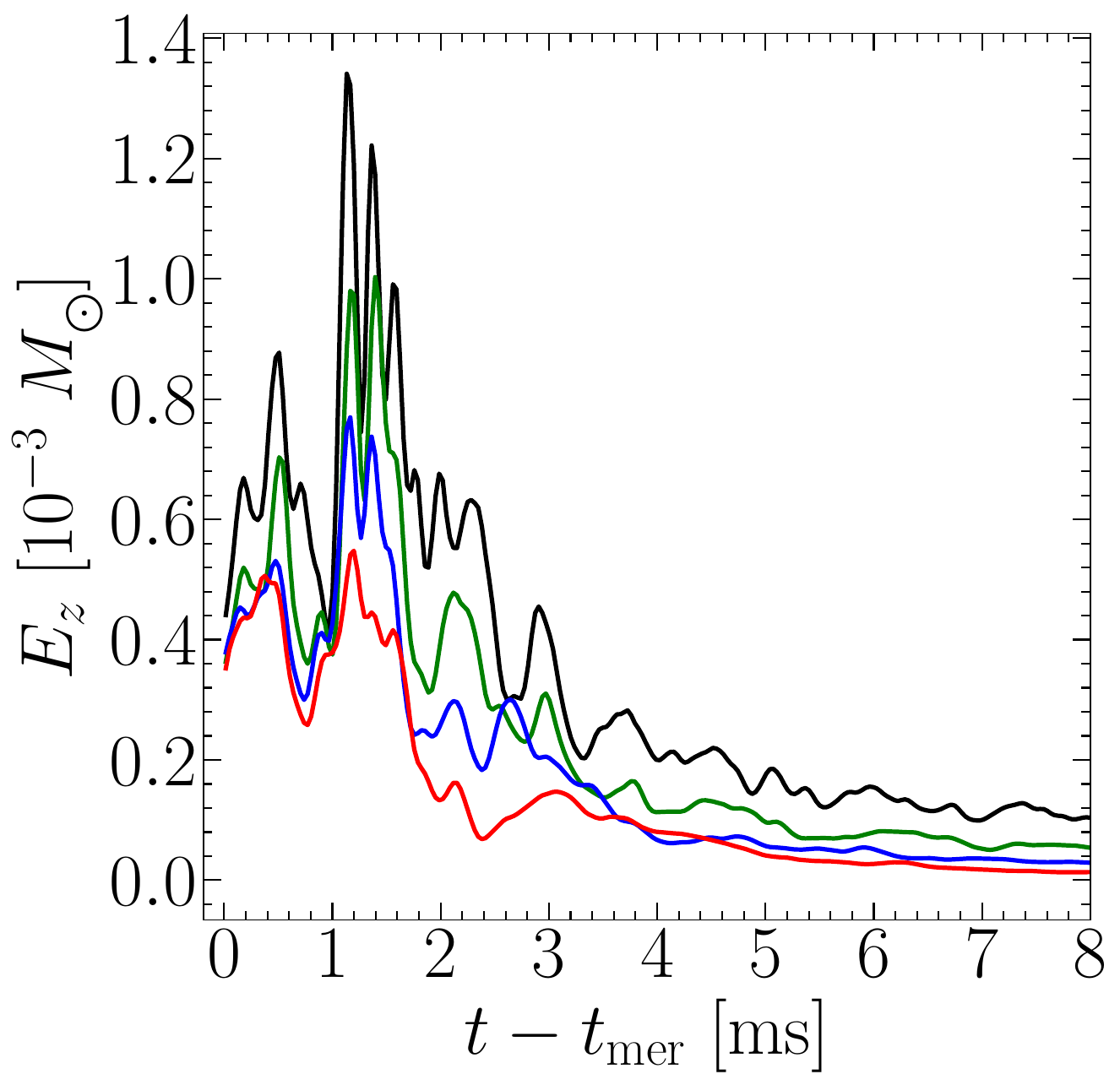}
\caption{\textit{Left}: Maximum rest-mass density in the postmerger phase for the
zero (black line), low (green line), medium (blue line) and high (red
line) viscosity case. The inset shows a zoom-in for the first five
milliseconds.
\textit{Middle}: Same as \textit{left} panel but for the total thermal energy
defined in Eq.~\eqref{eq:en_thermal}.
\textit{Right}: Same as \textit{left} panel but for the total
$z$-directed kinetic energy defined in Eq.~\eqref{eq:en_z}.
\label{fig:vis_thermal}
}
\end{figure*}

We find in the right panel of Fig.~\ref{fig:mig_shocks} that the inner
solid density contour line of the zero-viscosity case reaches larger
values in $x$ than the corresponding density contour line in the
medium-viscosity case. This indicates that the neutron star in the
medium-viscosity case has a less compact core which can be explained by
the additional thermal support observed in Fig.~\ref{fig:mig_central}. At
the same time a larger density gradient is present in the outer layers of
the medium-viscosity case which can be seen by comparing distances
between neighbouring density contour lines. As a result, the pressure
gradient in the outer layers of the medium-viscosity star is increased
which means that the neutron star surface and the location where the
shock forms have moved outwards in comparison to the zero-viscosity case.

Second, we observe that larger values for the bulk viscosity lead to less
energetic shock waves, especially when considering the second and third
shock wave. This can be seen in the left panel of Fig.~\ref{fig:mig_shocks} when
concentrating on the second shock wave for each value of the bulk
viscosity. Clearly, the velocities of the shock fronts decrease with
higher $\zeta$. In the high-viscosity case the shock front of
the second wave even stops propagating when it reaches $\sim 60\, \mathrm{km}$.
The reason for this behaviour is mainly related to the damping of the
neutron star oscillations due to viscosity. For the viscous cases matter
is still radially infalling and being shocked at the neutron star
surface. But due to the increased dissipative transfer of radial kinetic
energy to thermal energy in the neutron star a lower energy budget is
available to accelerate the shock. Hence, the shocks become weaker.

\subsection{Binary neutron star mergers}
\label{sec:binary}

\subsubsection{Simulation setup}

In this section we discuss the simulations of BNS mergers with a constant
and causal bulk-viscosity prescription. We use the initial data setup
presented in~\cite{Chabanov2023} which produces a long-lived HMNS
remnant. Similar to the numerical experiments with isolated stars, we
employ a hybrid EOS of the form
$p=p_{\mathrm{cold}}+\rho\epsilon_{\mathrm{th}}(\Gamma_{\mathrm{th}}-1)$
where the cold part is modeled by a cold, $\beta$-equilibrated slice of
the \texttt{TNTYST} EOS~\cite{Togashi2017}. We choose
$\Gamma_{\mathrm{th}}=1.7$ which corresponds to the optimal value found
in~\cite{Figura2020}. We simulate four different values for the bulk
viscosity $\zeta$ which is given by $\zeta_h \in
[\zeta_0,\zeta_0/2,\zeta_0/5,0]$ with
$\zeta_0=10^{30}~\mathrm{g}~\mathrm{cm}^{-1}~\mathrm{s}^{-1}$. The value
of $\zeta_0$ corresponds approximately to the highest bulk viscosity
observed in the~\cite{Most2022}. In accordance with the magnitude of
$\zeta$ the simulations are labeled as high, medium, low and zero
viscosity case. We choose $\rho_h \approx
4.52\times10^{14}~\mathrm{g}~\mathrm{cm}^{-3}$ and $\rho_l \approx
1.13\times10^{12}~\mathrm{g}~\mathrm{cm}^{-3}$. Besides, we have set
$\tau_h \approx 2.7\times10^{-4}\,\mathrm{ms}$.

All reported simulations are performed with a resolution of $\Delta x\sim
0.17~M_{\odot}\approx 260\,\mathrm{m}$ on the sixth refinement level. We
have also performed simulations with a lower resolution of $\Delta x\sim
0.25~M_{\odot}\approx 370\,\mathrm{m}$ for the zero, low and high
viscosity case. We observe qualitatively consistent behaviour such that
we only report the results of the high-resolution simulations.

\subsubsection{Thermal properties}

In this subsection we discuss the impact of bulk viscosity on thermal
properties of BNS mergers. We start by reporting the maximum rest-mass
density in the left panel of Fig.~\ref{fig:vis_thermal} and the thermal energy 
\begin{align}\label{eq:en_thermal}
E_{\mathrm{th}} := \int \rho
\epsilon W\, \sqrt{\gamma}\, dxdydz \,,
\end{align}
in the middle panel of Fig.~\ref{fig:vis_thermal}. The integral in
Eq.~\eqref{eq:en_thermal} is calculated over the whole simulation domain.

First, from the inset in the left panel of Fig.~\ref{fig:vis_thermal} we
observe that the strong density oscillations which occur during the first
$\sim 5~\mathrm{ms}$ after merger are damped efficiently in the cases
with strong bulk viscosities. This is not surprising as these
oscillations are related to expansion and compression cycles of the newly
formed HMNS. It is precisely the kinetic energy stored in fluid expansion
and compression which is dissipated through bulk viscosity.  Furthermore,
we observe that a larger bulk viscosity leads to a less dense HMNS core
at the end of our simulations at around $t-t_{\mathrm{mer}}\approx
24~\mathrm{ms}$. This effect is related to the increase in the kinetic
rotational energy which adds centrifugal support to matter located in
the densest regions of the HMNS (see Appendix \ref{sec:bnsrot} for
details). Note that in contrast to the constant bulk-viscosity
prescription applied in this work, a microphysical model leads to a
denser and more compact remnant with constant kinetic rotational energy
across different magnitudes of $\zeta$ \cite{Chabanov2023}.

Second, from the middle panel of Fig.~\ref{fig:vis_thermal} we find that
larger bulk viscosities lead to a larger overall thermal energy in the
first $\sim 5~\mathrm{ms}$ after merger. As observed before, during the
first $\sim 5~\mathrm{ms}$ after merger the damping of large density
oscillations is most efficient such that also the dissipation of kinetic
into thermal energy is most efficient. Thus, larger bulk viscosities lead
to larger thermal energies during that time. However, after the first
five milliseconds other effects need to be taken into account in order to
explain the evolution of $E_{\mathrm{th}}$. While the dissipation of
kinetic into thermal energy leads to an increase in thermal energy the
aforementioned decrease in central rest-mass density leads to a decrease
in thermal energy as the HMNS core is less dense as a result of the
additional centrifugal support. Both of these effects are competing on a
timescale larger than five milliseconds which leads to the non-monotonic
behaviour observed in $E_{\mathrm{th}}$ at the end of our simulations. In
particular, we see that the decrease in central density of the medium and
high-viscosity cases leads to \textit{lower} final thermal energies than
in the low viscosity case which has the largest final thermal energy. The
final thermal energy of the high-viscosity case is even below the
inviscid case.

Finally, we show the $z$-directed kinetic energy in the simulation domain
\begin{align}\label{eq:en_z}
E_{z} := \frac{1}{2} \int \rho
h v^{z}u_{z}\, \sqrt{\gamma}\, dxdydz \,,
\end{align}
in the right panel of Fig.~\ref{fig:vis_thermal}. In order to strengthen
the claim that during the first $\sim 5~\mathrm{ms}$ after the merger
most of the bulk-viscous energy dissipation is taking place we can make
use of $E_z$. This is illustrative as the kinetic energy in the
$z$-direction is generated purely by strong oscillations and shocks
coming from the collision of the binary and is therefore subject to
bulk-viscous damping. Here, we see very clearly that the behaviour is
monotonic and larger bulk viscosities lead to a smaller $E_z$.
Additionally, we also verify that most of the kinetic energy is
dissipated during the first $\sim 5~\mathrm{ms}$ after merger.

\begin{figure*}
\includegraphics[width=0.47\textwidth]{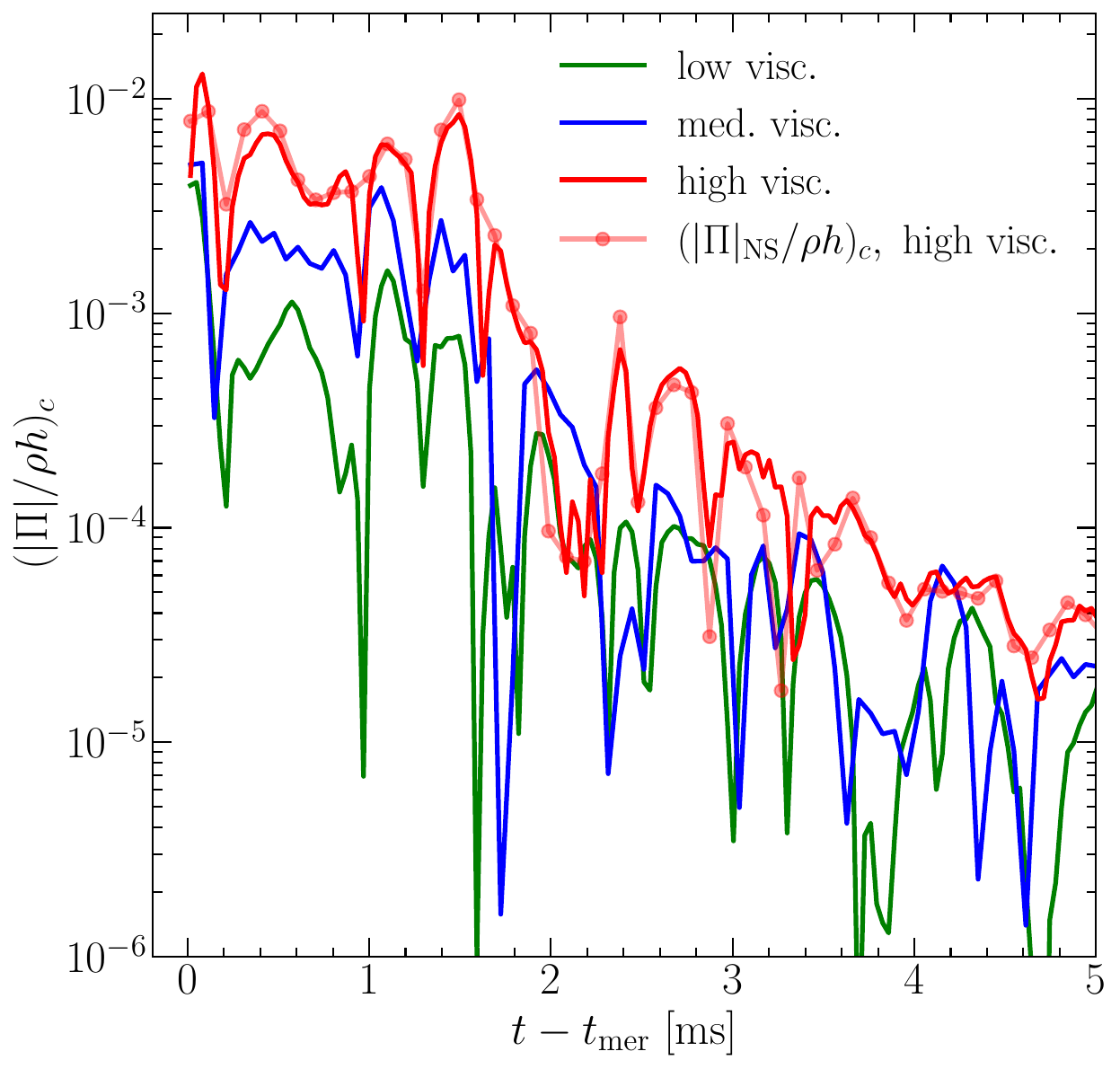}
\hskip 0.3cm
\includegraphics[width=0.47\textwidth]{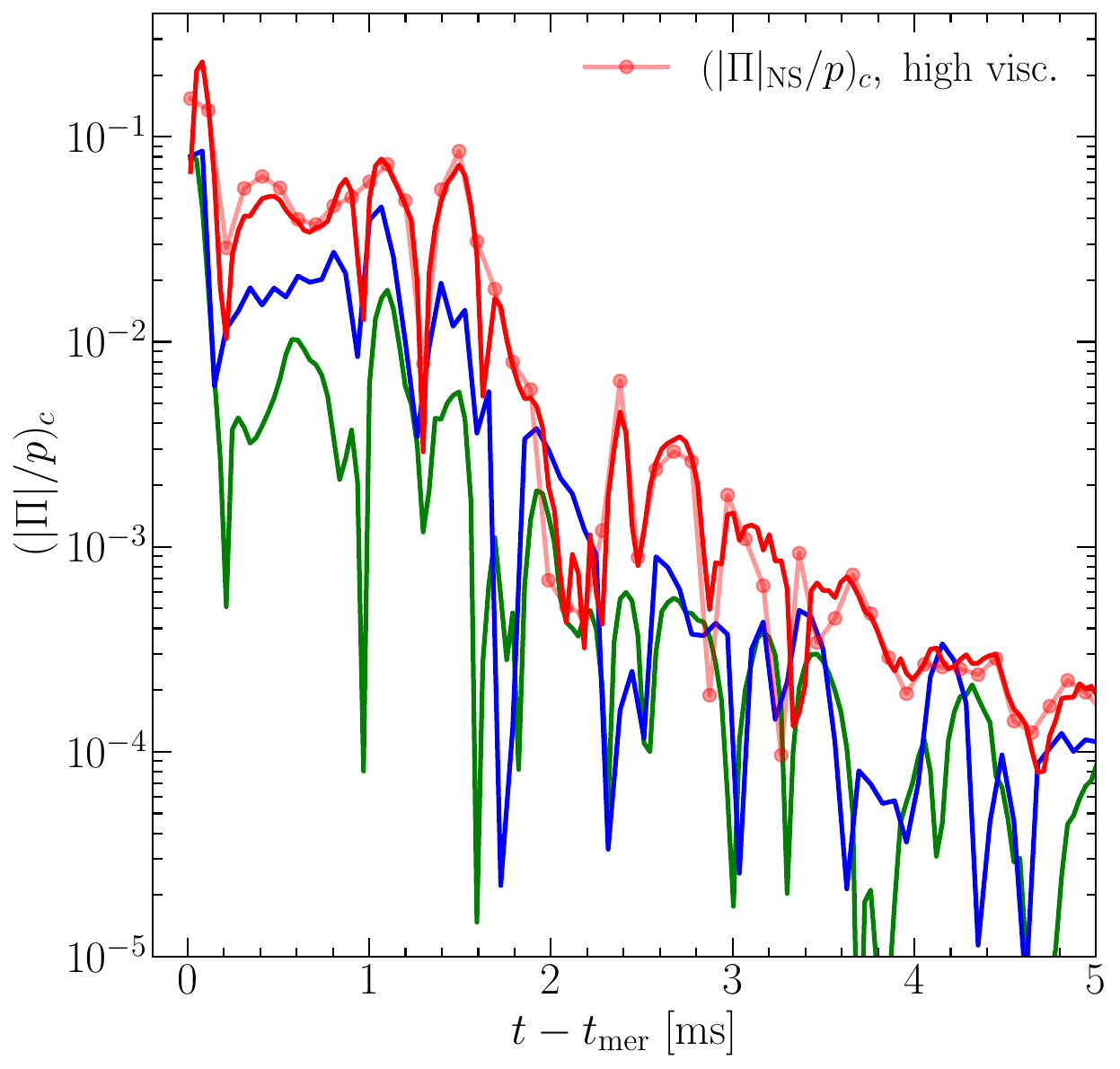}
\caption{\textit{Left}: Absolute inverse Reynolds number, \ie the ratio
  of the absolute bulk-viscous pressure over the enthalpy density, for
  the zero (black line), low (green line), medium (blue line) and high
  (red line) viscosity case. Transparent red circles denote the same
  quantity but calculated from the NS value for the high-viscosity case.
  \textit{Right}: Same as \textit{left} panel but for the ratio of the
  absolute bulk-viscous pressure over the EOS pressure.
\label{fig:bin_central}
}
\end{figure*}

Furthermore, we can also evaluate ``how viscous'' the fluid is or in
other words, how large the deviations from thermodynamic equilibrium are
by measuring the relativistic Reynolds number $\mathcal{R}$ defined in
Eq.~\eqref{eq:reynolds}. We present the quantity $|\mathcal{R}^{-1}|$
measured in the center of the grid for the first $\sim 5~\mathrm{ms}$
after merger in the left panel of Fig.~\ref{fig:bin_central} and
$|\Pi|/p$ measured in the center of the grid for the same time interval
in the right panel. We present timeseries for all simulations and
additionally show the quantities $(|\Pi|_{\mathrm{NS}}/\rho h)_c$ and
$(|\Pi|_{\mathrm{NS}}/p)_c$ denoted by red transparent lines and circles
in both panels for the high-viscosity case, see also
Eq.~\eqref{eq:nsvalue}.

Both panels show qualitatively the same behaviour: Larger
bulk viscosities lead to a larger inverse Reynolds number as well as
larger viscous contributions to the total pressure. It is interesting to
note that for the high-viscosity case the inverse Reynolds number reaches
values above $\sim 1~\%$ which corresponds to a bulk-viscous pressure that
is $\sim 20~\%$ of the equilibrium pressure. This happens right after
merger when the strongest density oscillations are present. Afterwards,
both quantities decrease but remain of this order until
$t-t_{\mathrm{mer}}\approx 1.5~\mathrm{ms}$.

Then, both quantities drop sharply by approximately two orders of
magnitude and, after further decrease, reach values of around
$(|\Pi|/\rho h)_c\approx 10^{-5}$ and $(|\Pi|/p)_c\approx10^{-4}$ at
$t-t_{\mathrm{mer}}\approx 5~\mathrm{ms}$. This provides further evidence
that bulk-viscous dissipation works very efficiently only in the first
five milliseconds after merger; Figure~\ref{fig:bin_central} even
suggests that during the first $1.5~\mathrm{ms}$ after merger
bulk-viscous dissipation is having the biggest impact on the merger.
Additionally, similar to the results presented for the migration test in
Fig.~\ref{fig:mig_central}, we observe good agreement between the true
bulk-viscous pressure $\Pi$ and $\Pi\ns$ for all viscosity cases (note
that NS values are presented only for the high-viscosity case to avoid
overcrowding the figure).

\begin{figure*}
\includegraphics[width=0.99\textwidth]{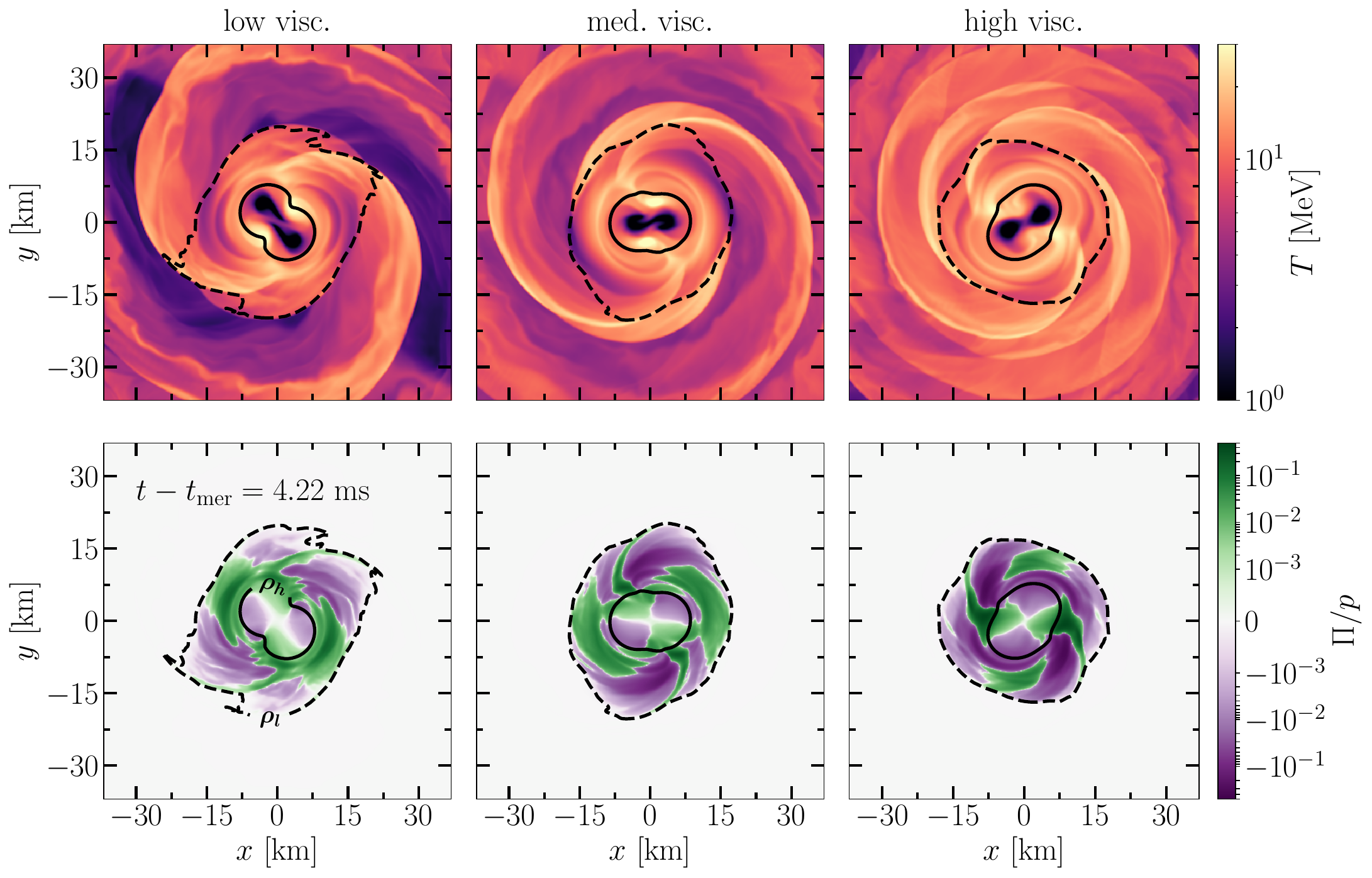}
\caption{Cross-sections of the temperature (top row) and of the ratio of
  the bulk-viscous pressure over the EOS pressure (bottom row) in the
  $(x,y)$ plane at $z=0$. The snapshots are taken at
  $t-t_{\mathrm{mer}}=4.22~\mathrm{ms}$. Shown are the low viscosity
  (left column), medium viscosity (middle column) and high-viscosity
  (right column) cases. The solid line denotes a density contour at $\rho
  = \rho_h$ while the dashed line denotes a density contour at $\rho =
  \rho_l$.
\label{fig:bin_temp_pip}
}
\end{figure*}

The measurement of the relativistic Reynolds number enables us to compare
out-of-equilibrium contributions in BNS mergers to the viscous contributions
in other relativistic fluids, namely the quark-gluon plasma (QGP)
encountered in heavy-ion collisions (HICs). Following the results
presented in~\cite{Most2022_a}, we first observe that the
order-of-magnitude estimates obtained in that study through post-process
calculations agree with our findings for the highest viscosity case. This
is not surprising as~\cite{Most2022_a} presents the maximum inverse
Reynolds number over the whole domain which is likely found where the
bulk viscosity $\zeta$ reaches its maximum. Since our high-viscosity case
makes use of a constant value for $\zeta$ which corresponds to the
highest value found in~\cite{Most2022_a}, it is natural to expect similar
inverse Reynolds numbers if the underlying fluid motion does not vary
significantly. 

Second, as observed in~\cite{Most2022_a} we find that the
inverse Reynolds number in HICs is approximately one order of magnitude
higher than in BNS mergers suggesting that the QGP is further away from
thermodynamic equilibrium than the high-density matter encountered in a
HMNS shortly after merger. 

Next, we report in Fig.~\ref{fig:bin_temp_pip} sections of the
temperature distribution (top row) and the distribution of the quantity
$\Pi/p$ (bottom row) in the equatorial plane at
$t-t_{\mathrm{mer}}\approx 4.22~\mathrm{ms}$ for the low (left column),
medium (middle column) and high viscosity (right column) cases. The
solid line denotes a density contour at $\rho = \rho_h$ while the dashed line
denotes a density contour at $\rho = \rho_l$. During this time the
strong damping of density oscillations and thus the dissipation of
oscillatory kinetic energy into thermal energy has almost ended, as
discussed previously. However, the HMNS has not settled into an
approximately axisymmetric state yet where the rotational energy density
has been redistributed. Thus, during this time the thermal energy is not
yet affected by the indirect influence of matter redistribution due to
centrifugal forces.

This also means that we should be able to observe higher temperatures for
higher bulk viscosities during this time because all configurations have
a similar compactness as can be observed in Fig.~\ref{fig:vis_thermal},
where all configurations have a comparable maximum rest-mass density at
$t-t_{\mathrm{mer}}\approx 4.22~\mathrm{ms}$. Indeed, we observe that the
temperature distributions shown in the top row of
Fig.~\ref{fig:bin_temp_pip} look more homogeneous and feature less
``cold'' material, \ie $T\lesssim5~\mathrm{MeV}$, if the bulk viscosity
is large. Two effects are important in order to understand this
behaviour. First, as already mentioned, higher bulk viscosities lead to
more dissipation of kinetic into thermal energy which leads to an
increase in temperature if the fluid is not expanding simultaneously
(adiabatic cooling might dominate the temperature evolution at a later
stage). Second, the decrease in kinetic energy is affecting the
redistribution and ejection of loosely bound matter through shock waves.
This means that in addition to the dissipative heating due to bulk
viscosity the HMNS is also less efficient in transporting heated matter
through shock waves. Both of these effects lead to the relatively
homogeneous distribution of temperature in the outer layers of the HMNS
for the high-viscosity case shown in the top right panel of
Fig.~\ref{fig:bin_temp_pip}. 

In addition, the lower row shows how the bulk-viscous pressure is
distributed. First, given that $\rho_l$ marks the value of the rest-mass
density below which we set $\zeta=\zeta_l=0$, \eg see
Eq.~\eqref{eq:zeta_bound}, we observe that the bulk-viscous pressure
$\Pi$ becomes vanishingly small for densities below $\rho_l$.
Furthermore, we observe that $|\Pi|/p$ reaches its maximum value in the
transition zone $\rho_l < \rho < \rho_h$, \ie between the dashed and the
solid line. The largest values of $|\Pi|$ are reached in regions with
densities higher than $\rho_h$ because these regions have the largest
oscillations and bulk viscosities. Nevertheless, the largest values of
$|\Pi|/p$ are sensitive to the rapid decrease in pressure in the outer
layers of the HMNS which is the reason for $|\Pi|/p$ reaching its maximum
for $\rho < \rho_h$. Note that, in contrast to the constant
bulk-viscosity prescription applied in this work, a microphysical model
leads to distributions of $|\Pi|/p$ which are highly sensitive to the
temperatures achieved in the HMNS \cite{Chabanov2023}.

We also observe that even though the distributions of $\Pi/p$ are subject
to different density oscillations, overall, larger bulk viscosities lead
to larger bulk-viscous pressures as one would intuitively expect. We
recall that this was not the case for the viscous migration test reported
in Fig.~\ref{fig:mig_central} as the high-viscosity case was extremely
efficient in the damping of density oscillations leading to non-monotonic
behaviour in the late evolution. 

\begin{figure}
\includegraphics[width=0.49\textwidth]{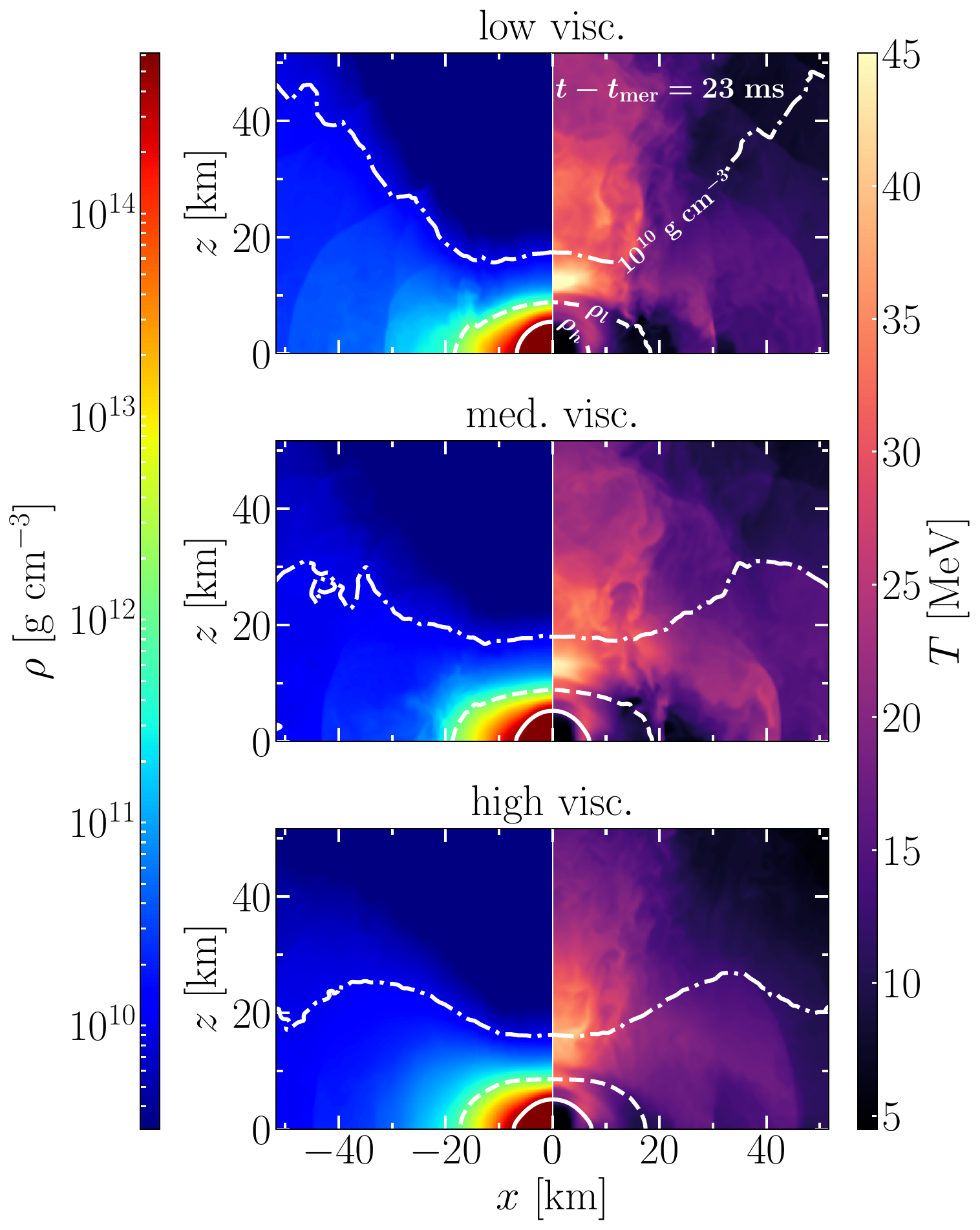}
\caption{\textit{Left half}: Cross-sections of the rest-mass density in
  the $(x,z)$ plane at $y=0$. The snapshots are taken at
  $t-t_{\mathrm{mer}}=23~\mathrm{ms}$. Shown is the low viscosity case in
  the top panel, the medium-viscosity case in the middle panel and the
  high-viscosity case in the bottom panel. The solid line denotes a
  density contour at $\rho = \rho_h$, the dashed line denotes a density
  contour at $\rho = \rho_l$ and the dash-dotted line denotes a low
  density contour at $\rho = 10^{10}~\mathrm{g}~\mathrm{cm}^{-3}$.
  \textit{Right half}: Same as \textit{left half} but for the
  temperature.
\label{fig:rho_temp_xz}
}
\end{figure}

Last but not least, the distributions of $\Pi/p$ resemble quadrupolar
structures for $\rho_h < \rho$ for all cases. This can be explained by
considering the $m=2$ bar-mode deformation of the HMNS and its
interaction with the surrounding matter. It is useful to recall that
$\Pi \sim -\zeta \Theta$ in our simulations. This means that regions
with $\Pi < 0$ are expanding while regions with $\Pi > 0$ are
compressing. First, we observe that one of the intersections between
negative and positive bulk-viscous pressure is connecting the two
temperature ``hot spots''~\cite{Hanauske2016} which also correspond to
the two largest eddies that developed during the merger. Then, the other
intersection separating negative and positive $\Pi$ regions is found
approximately orthogonal to the first intersection connecting the hot
spots. As the HMNS is rotating counter-clockwise in
Fig.~\ref{fig:bin_temp_pip} we observe that the regions of the HMNS which
are located immediately above either of the two hot spots are always
compressing (positive $\Pi$) while the regions below either of the two
hot spots are expanding (negative $\Pi$). As the two eddies located at
the two temperature hot spots have the same orientation as the rotation
axis of the HMNS, \ie both rotate counter-clockwise, matter above the hot
spots must have a positive $\Pi$ and matter below a negative $\Pi$. The
reason for this is that the rotating motions of the eddies compress
matter in front of them because of the drag towards the HMNS which means
that its density is increasing. Correspondingly, matter behind the eddies
expands because of the drag away from the HMNS which means that its
density is decreasing.

Finally, in Fig.~\ref{fig:rho_temp_xz} we discuss sections of the
rest-mass density distribution (left half panels) and the distribution of
the temperature (right half panels) in the $(xz)$ plane ($y=0$) at
$t-t_{\mathrm{mer}}\approx 23~\mathrm{ms}$ for the low (top panel),
medium (middle panel) and high viscosity (bottom panel) cases.
The solid line denotes a density contour at
$\rho = \rho_h$,
the dashed line denotes a density contour at $\rho = \rho_l$ and the dash-dotted line
denotes a low density contour at $\rho =
10^{10}~\mathrm{g}~\mathrm{cm}^{-3}$. This figure helps us to understand the impact of
large bulk viscosities on the overall structure of the long-lived HMNS
and torus when a quasi-stationary state is reached. First, we observe
that the torus has a smaller extent, if the bulk viscosity is increased.
Also, the HMNS has a less oblate shape with larger bulk viscosities which
is related to the less efficient angular momentum transport to the outer
layers of the remnant. In general, large bulk viscosities prevent neutron
star matter from reaching large radii which leads to an overall more
compact envelope and an higher concentration of angular momentum at
smaller radii. 
 
Second, focussing on the temperature distributions, we find that the low
density funnel above the HMNS has overall lower temperatures with larger
bulk viscosities. At this point it is important to remark that large
temperatures in the funnel region are most likely the consequence of
using a hybrid EOS with an ideal gas law because a comparable study using
fully tabulated hot EOSs did not show this behaviour ~\cite{Radice2017}.
However, since we are interested in learning about the qualitative impact
that large constant bulk viscosities have in the postmerger phase, it is
still interesting to compare the temperature distributions for the
different cases in order to understand how bulk viscosity can possibly
impact the funnel region. In our case, large temperatures in the funnel
region stem from bound shock-heated material which accumulates above the
HMNS. As bulk viscosity tends to damp violent nonlinear density
oscillations of the HMNS, which are a primary source of shock waves,
larger bulk viscosities tend to produce weaker shock waves and therefore
less efficient shock-heating in the funnel region.

\begin{figure}
\includegraphics[width=0.49\textwidth]{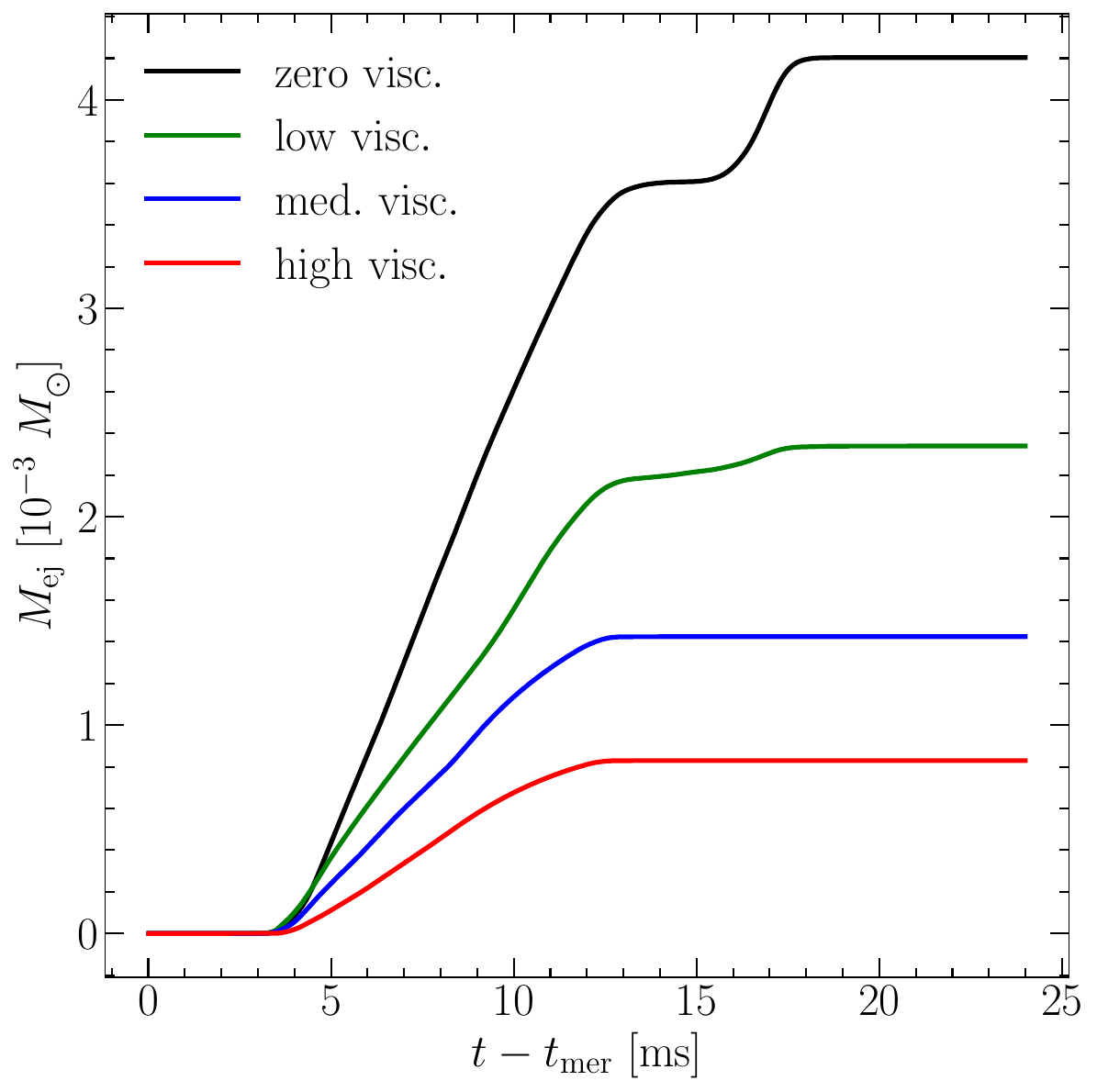}
\caption{Cumulative ejected mass from the
Bernoulli criterion for the
zero (black line), low (green line), medium (blue line) and high (red
line) viscosity case. The detector is placed at $\sim 517~\mathrm{km}$
for all viscosity cases.
\label{fig:outflow_tot}
}
\end{figure}

\subsubsection{Dynamical mass ejection}

\begin{figure}
\includegraphics[width=0.49\textwidth]{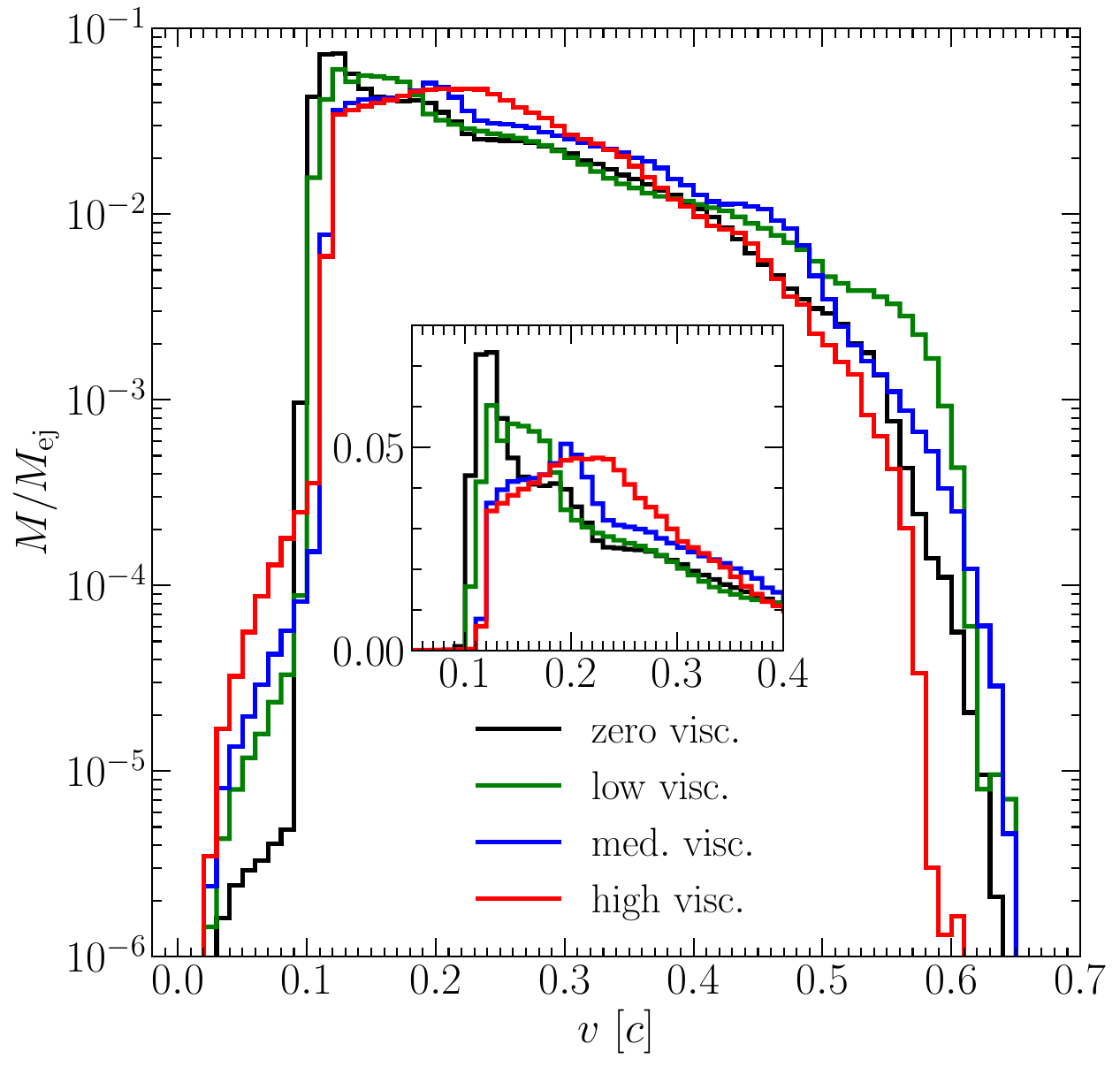}
\caption{Velocity distribution of the ejected mass with 
$v=\sqrt{v_{i}v^{i}}$ at the end of our simulations for the
zero (black line), low (green line), medium (blue line) and high (red
line) viscosity case. The inset
shows a zoom-in around the mean of the distribution. The detector is placed at $\sim 517~\mathrm{km}$
for all viscosity cases.
\label{fig:outflow_dist}
}
\end{figure}

In this subsection we discuss the impact of large bulk viscosities on the
dynamical mass ejection in our simulations. We start by showing the total
unbound material passing through a detector at $\sim 517~\mathrm{km}$ and
calculated by using the Bernoulli criterion, \ie material is classified
as being unbound if $-hu_t-h_{\infty}>0$ with $h_{\infty}$ being the
minimum specific enthalpy of the cold EOS table employed, see also
~\cite{Fujibayashi2023}. Figure \ref{fig:outflow_tot} shows the
cumulative unbound mass as a function of $t-t_{\mathrm{mer}}$ for all
viscosity cases. We observe a clear impact of bulk viscosity on the total
unbound material. Large bulk viscosities suppress dynamical mass
ejection. For the high-viscosity case the total ejecta mass is only $\sim
20~\%$ of the mass measured for the zero-viscosity case. 

\begin{figure*}
\includegraphics[width=0.99\textwidth]{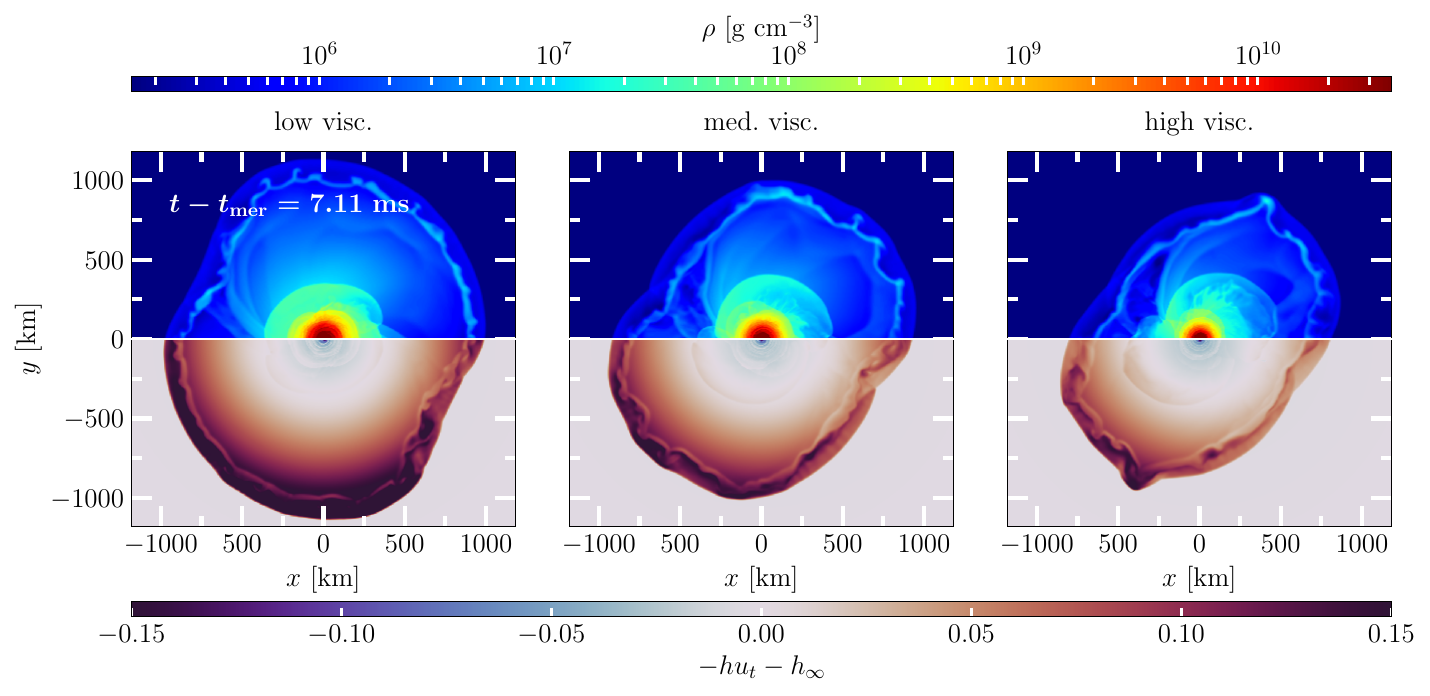}
\caption{\textit{Top half}: Cross-sections of the rest-mass density
in the $(x,y)$ plane at $z=0$. The snapshots are taken at
$t-t_{\mathrm{mer}}=7.11~\mathrm{ms}$. Shown is the low viscosity case in
the left panel, the medium viscosity case in the middle panel and the
high viscosity case in the right panel. \textit{Bottom half}: Same as
\textit{top half} but for the Bernoulli criterion. 
\label{fig:bin_bernoulli}
}
\end{figure*}

Figure~\ref{fig:outflow_dist} shows the velocity distribution of the
total unbound material at the end of the simulations for all viscosity
cases. The impact of bulk viscosity on the velocity distributions is less
severe than its impact on the ejected mass. We observe that large bulk
viscosities tend to suppress ejecta with velocities below $\sim 0.2$ and
above $\sim 0.5$ compared to the less viscous cases. Therefore, as can be
seen from the inset in Fig.~\ref{fig:outflow_dist}, the median of the
distribution tends to shift to higher velocities. We suspect that the
slow ejecta, more precisely, ejecta with $v \lesssim 0.2$, observed in
the zero and low viscosity cases stems from marginally unbound material
which is accelerated by shock waves which pass through it. Large
viscosities would naturally suppress this type of ejecta due to less
efficient shock-heating, see also the discussion of
Figs.~\ref{fig:bin_temp_pip} and \ref{fig:rho_temp_xz}. As a consequence,
the distribution of the ejecta in $v$ shifts to higher velocities as only
the marginally unbound material with low velocities is affected by this
mechanism. The suppression of the \textit{fast} ejecta component is the
direct result of strong viscous damping of early density oscillations.
The fast ejecta component originates from the first bounces of the two
neutron star cores. As bulk viscosity tends to damp these violent
collisions most efficiently, the energy reservoir powering the ejection
of fast matter is dissipated and leads to a reduction of fast unbound
matter.

\begin{figure}
\includegraphics[width=0.49\textwidth]{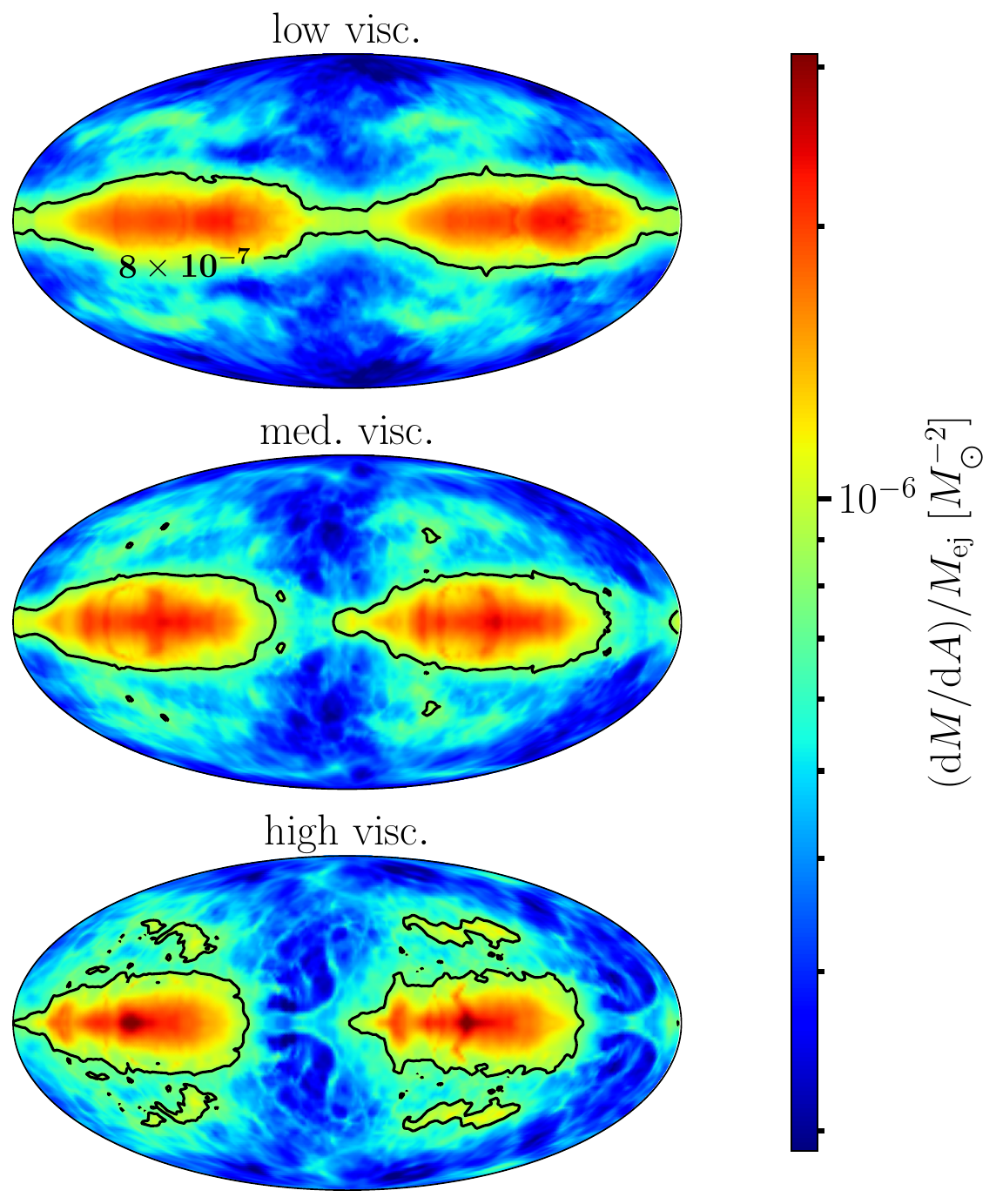}
\caption{Mollweide projections of the surface density as calculated from
  the cumulative ejected mass per solid angle measured by using the
  Bernoulli criterion at the end of our simulations. The values are
  normalized by the corresponding total mass of the ejecta. The detector
  surface is placed at $\sim 517~\mathrm{km}$. Shown is the low
  viscosity case in the top panel, the medium-viscosity case in the
  middle panel and the high-viscosity case in the bottom panel.  Solid
  lines denote contour lines at $8\times10^{-7}~M_{\odot}^{-2}$.
\label{fig:outflow_proj}
}
\end{figure}

Next, we will discuss the geometric distribution of the dynamical ejecta.
Figure \ref{fig:bin_bernoulli} shows the rest-mass density (top half
panels) and the quantity $-hu_t-h_{\infty}$ (bottom half panels) at
$t-t_{\mathrm{mer}}\approx7.11~\mathrm{ms}$ for all configurations in the
equatorial plane. We observe that an increase in bulk viscosity leads to
a more asymmetric shape of the outermost ejecta front which is at the
same time a major component of the overall dynamical ejecta. This means
that large bulk viscosities tend to select a preferred direction of the
dynamical ejecta in the $\phi$-direction. In our case, mass ejection
tends to be increased for the angles $\phi\approx \pi/4$ and $\phi\approx
5\pi/4$. This finding is further supported by the Mollweide projections
of the normalized surface density as calculated from the cumulative
ejected mass per solid angle presented in Fig.~\ref{fig:outflow_proj}.
Solid lines denote contour lines at $8\times10^{-7}~M^{-2}_{\odot}$. For
the medium and high-viscosity case we observe pockets of low mass
ejection in the equatorial plane with less than
$8\times10^{-7}~M^{-2}_{\odot}$. On the other hand, we also observe that
the medium and high-viscosity cases tend to reach higher densities inside
the contour lines.

This effect can easily be explained by considering that the dynamical
ejecta consists of a tidally ejected component and a shock-heated
component~\cite{ShibataRev19}. The tidally ejected component is weakly
affected by bulk viscosity in our simulations as we set $\zeta=0$ in low
density matter which is more likely to get unbound through tidal forces.
In contrast, as already explained above together with
Fig.~\ref{fig:rho_temp_xz}, strong bulk viscosities reduce shock-heating.
Hence, the first bounce of the two neutron star cores powers
significantly more matter ejection than all subsequent bounces, if the
bulk viscosity is large. Therefore, matter is ejected primarily in the
direction of the first bounce for the high-viscosity case, while for the
cases with a lower viscosity matter, ejection from the subsequent core
bounces tends to be more symmetric along the $\phi$-direction.
Furthermore, from Fig.~\ref{fig:outflow_proj} we observe only a weak
influence of bulk viscosity on the surface density distribution in the
$\theta$-direction. In the contrast to the $\phi$-direction, the surface
density in the high-viscosity case is more uniform along $\theta$ for a
given $\phi$ than in the low viscosity case. Thus, while large bulk
viscosities increase the degree of anisotropy of the outflow in the
$\phi$-direction, they also show a weak tendency to decrease anisotropy
in the $\theta$-direction.

\begin{figure}
\includegraphics[width=0.49\textwidth]{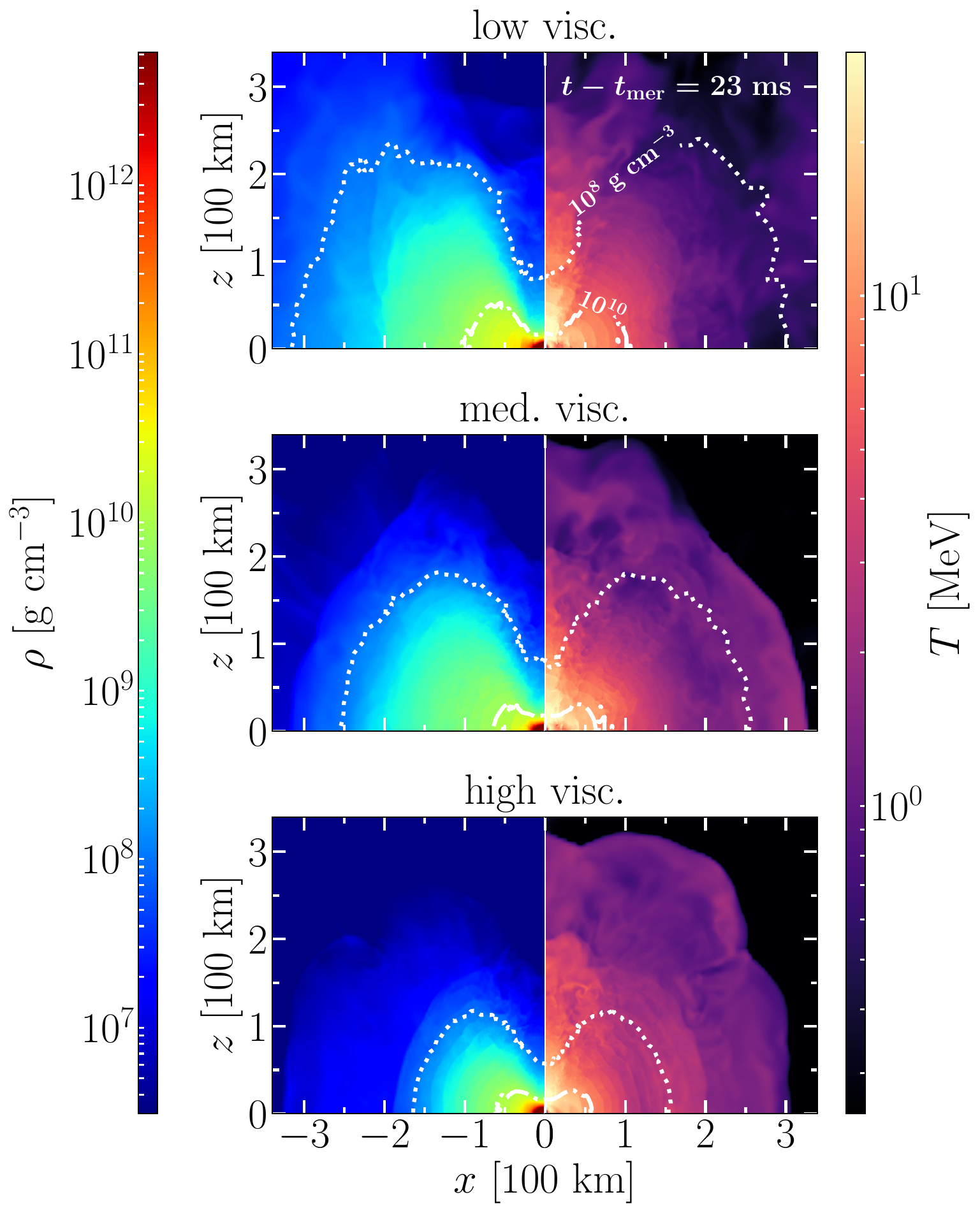}
\caption{Same as Fig.~\ref{fig:rho_temp_xz} but for a much larger domain focussing
on the envelope of the remnant. Note that we present an additional low
density contour at $\rho = 10^{8}~\mathrm{g}~\mathrm{cm}^{-3}$ by using a
dotted line and that the colorbars have a different scale.  
\label{fig:rho_temp_xz_large}
}
\end{figure}

To conclude this subsection, we present in
Fig.~\ref{fig:rho_temp_xz_large} the same quantities as in
Fig.~\ref{fig:rho_temp_xz} focussing now, however, on the distributions
at large distances from the HMNS. Additionally, we present another low
density contour at $\rho = 10^{8}~\mathrm{g}~\mathrm{cm}^{-3}$ using a
dotted lines in order to visualize the envelope or torus. We find that
large bulk viscosities lead to a significantly more compact envelope or
torus as can be seen from the size of the different areas enclosed by the
dotted line. We attribute this observation to the less efficient
transport of matter to large distances which is a direct effect of large
bulk viscosities. The temperature distributions at large distances are
comparable for all cases which indicates that the impact of bulk
viscosity on the thermal properties of matter located in the torus is
weak.

\section{Conclusions}
\label{sec:conc}

We present a comprehensive report about the impact of a constant bulk
viscosity on BNS mergers using numerical simulations where dissipative
effects are included self-consistently through the causal and
second-order formulation by MIS. In the interest of exploring the
most elementary scenario, we make use of a particularly simplified
microphysical model for the bulk viscosity where it assumes a constant
value within the neutron stars and is set to zero below a certain
threshold density. This approach is simpler than the one carried out in
Ref.~\cite{Chabanov2023}, where the transport coefficients arise
consistently from microphysical arguments leading to complex functions
of density and temperature. At the same time, the simpler approach
presented here has the important advantage of being independent of the
EOS and thus more suitable for serving as a numerical testbed.

First, we present a detailed description of the numerical methods used in
\cite{Chabanov2023} and of the numerical tests performed in order to
verify our implementation. Most notably, we employ a modified version of
the inversion algorithm proposed in~\cite{Galeazzi2013}, where several
limiting procedures ensure causality of the evolution, as well as
existence and uniqueness properties of the root-finding function used in
the conversion between conservative and primitive variables.

As a first test case we consider the measurement of numerical viscosity
and calculate the damping time of the radial fundamental eigenmode of an
isolated TOV solution with bulk viscosity. By evolving stars with
different physical bulk viscosities and different resolutions we were
able to observe that the measured bulk viscosity asymptotes to a constant
value for high resolutions. The measured asymptotic value differs at most
by $\lesssim 35~\%$ from the physical input viscosity which is a
reasonable agreement considering the strong approximations used in the
measurement procedure. Additionally, we find that the convergence order
of our code is strongly affected by the the relative importance of
discretization errors originating from the ill-balanced surface or the
smooth interior. In a representative setup we obtained numerical
viscosities $\lesssim
10^{26}~\mathrm{g}~\mathrm{cm}^{-1}~\mathrm{s}^{-1}$ which is a promising
finding considering that realistic viscosities reach values of $\gtrsim
10^{27}~\mathrm{g}~\mathrm{cm}^{-1}~\mathrm{s}^{-1}$
~\cite{Schmitt2018,Alford2020} in isolated stars and BNS mergers.

As a second test case we simulate the violent migration of unstable
neutron stars to the stable branch. In contrast to the first test case,
this scenario tests the implemented equations in a nonlinear regime. We
observe that dissipative heating leads to a decrease of the central
rest-mass density because the contribution of the thermal pressure in the
neutron star core increases. The contribution can reach values of up to
$\sim 10~\%$ of the total pressure for the highest viscosity
case. Additionally, this test case allows us to provide evidence that our
scheme is able to handle the transition between a viscous neutron star
and an inviscid low density exterior. This is an important ability as it
allows us to study the impact of bulk viscosity on mass ejection in BNS
merger simulations. We show that shock waves which originate at the
neutron star surface located in viscous matter propagate without any
numerical disturbance through the transition region between viscous and
inviscid matter. This allows us to observe that higher viscosities lead
to weaker shock waves in the second and third bounce during migration. It
is due to the efficient dissipation of kinetic energy into heat that a
lower amount of energy is transferred to the shock wave for higher bulk
viscosities.

Finally, we explore BNS simulations with a constant bulk
viscosity. We report on thermal and structural properties of the HMNS
remnant, as well as the impact of bulk viscosity on dynamical mass
ejection. We find that for large viscosities, the increase of the total
thermal energy due to dissipative heating is counteracted by the lower
density of the HMNS core. These lower densities are the result of the
additional centrifugal support in the HMNS core observed. It is
important to remark that when dropping the assumption of constant bulk
viscosity and adopting instead a microphysical model for the latter,
leads to a denser and more compact remnant due to the strong
temperature dependence of the microphysical bulk viscosity
\cite{Chabanov2023}.

We also observe a more uniform temperature distribution and less
efficient shock-heating in the HMNS remnant. We attribute this behaviour
to the dissipation of kinetic energy into heat and, as a result, a less
efficient kinetic-energy transfer from the fast-rotating deformed HMNS
core to the matter making up its envelope. Furthermore, we measure
inverse Reynolds numbers on the order of $\sim 1~\%$ in the center of the
HMNS directly after the merger for the highest viscosity considered. This
corresponds to bulk-viscous pressures, which are approximately $\sim
20~\%$ of the total EOS pressure. Overall, the remnant of a viscous BNS
merger has a significantly smaller torus and at the same time a less
dense core.

Finally, the dynamical mass ejection from viscous BNS mergers is in
agreement with the overall results presented so far. We find that the
dynamically ejected mass in our simulations is suppressed by a factor of
approximately five for the high-viscosity case when compared to the
inviscid case. Again, this is the result of efficient dissipation of
kinetic energy which makes it hard to unbind mass. Interestingly, we also
find that the distribution of ejected matter along the azimuthal
direction becomes more anisotropic with increasing viscosities.
We attribute this behaviour to the fact that for large viscosities most
of the unbound material stems from the first collision-and-expansion
cycle of the two stars. Thus, matter is ejected in a preferred direction.

\begin{acknowledgments}
  It is a pleasure to thank M. Alford, M. Hanauske, E. Most, and
  K. Schwenzer for useful comments and discussions. Partial funding comes
  from the GSI Helmholtzzentrum f\"ur Schwerionenforschung, Darmstadt as
  part of the strategic R\&D collaboration with Goethe University
  Frankfurt, from the State of Hesse within the Research Cluster ELEMENTS
  (Project ID 500/10.006), by the ERC Advanced Grant ``JETSET: Launching,
  propagation and emission of relativistic jets from binary mergers and
  across mass scales'' (Grant No.  884631) and the Deutsche
  Forschungsgemeinschaft (DFG, German Research Foundation) through the
  CRC-TR 211 ``Strong-interaction matter under extreme conditions''--
  project number 315477589 -- TRR 211. LR acknowledges the Walter Greiner
  Gesellschaft zur F\"orderung der physikalischen Grundlagenforschung
  e.V. through the Carl W. Fueck Laureatus Chair. The simulations were
  performed on HPE Apollo HAWK at the High Performance Computing Center
  Stuttgart (HLRS) under the grant BNSMIC.
\end{acknowledgments}

\appendix
\section{Measurement procedure}
\label{sec:measurement_proc}

\begin{figure*}
\includegraphics[width=0.47\textwidth]{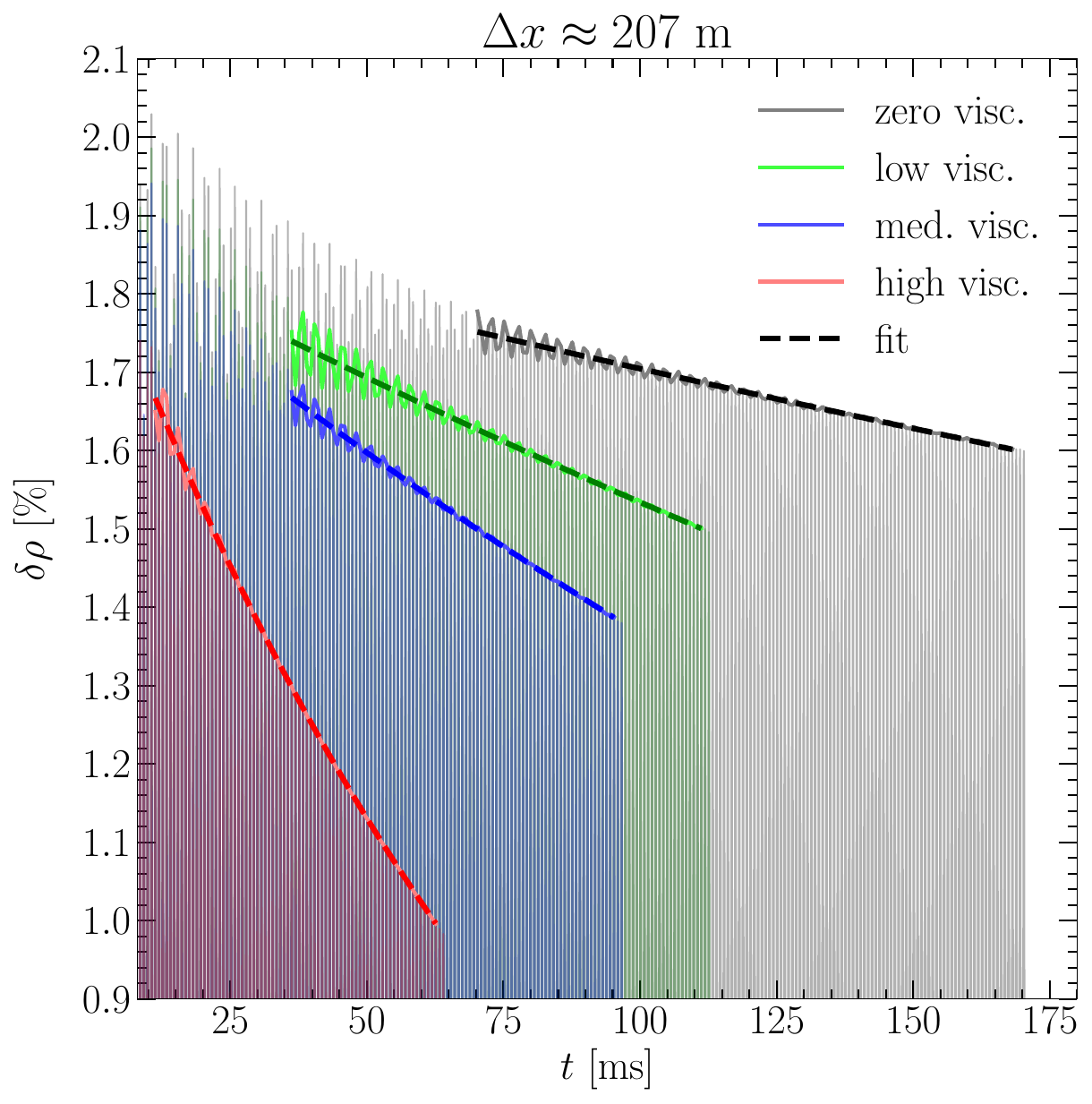}
\hskip 0.5cm
\includegraphics[width=0.47\textwidth]{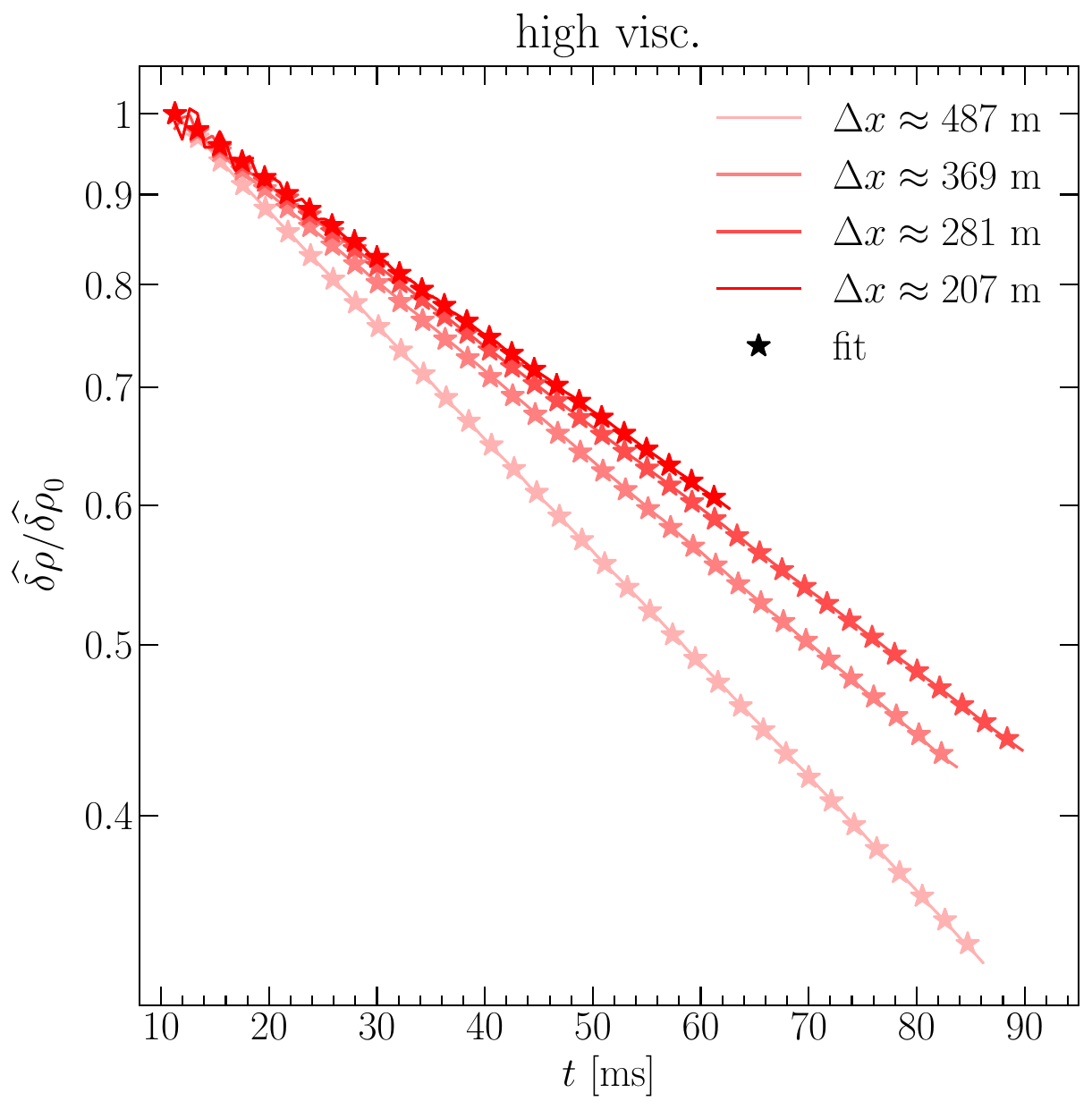}
\caption{\textit{Left}: Shown in solid thin lines is the timeseries of
the normalized central density for the zero (black line), low (green
line), medium (blue line) and high (red line) viscosity case, where the
density drift has been removed from the data. Shown in solid thick lines
is the corresponding envelope of $\delta \rho$ estimated by using the
local maxima of the signal. Note that the envelopes start at $t \neq 0$
indicating that data before the starting point has been discarded.
Corresponding dashed lines denote fits to the envelope of $\delta \rho$.
Overall, we show only data for the highest resolution, \ie $\Delta x
\approx 207~\mathrm{m}$.  \textit{Right}: Solid lines show the timeseries
of the local maxima of $\delta \rho$ normalized by its first value. We
show only data for the high-viscosity case but for all employed numerical
resolutions by using different shades of red. Higher resolutions are
represented by a darker shading. The star symbols denote fits to
$\widehat{\delta \rho}/\widehat{\delta \rho}_0$.
\label{fig:osc_amp}
\label{fig:osc_evol}}
\end{figure*}

Following~\cite{CerdaDuran2010} and~\cite{Cutler1990} the rate of change
of the kinetic energy of a weakly pulsating star, \ie the magnitude
of the pulsation is small such that linear perturbation theory is
applicable, is given by
\begin{align}
\frac{\mathrm{d}E}{\mathrm{d}t}=-4\pi\int_0^R\mathrm{d}r r^2\sqrt{-g}\zeta|\Theta|^2\,,
\end{align}
where $g$ is the determinant of the spacetime metric $g_{\mu\nu}$ and $R$
is the radius of the neutron star. Then, if the pulsation energy $E$ is
known, the damping time of density and velocity
perturbations $\overline{\tau}$ can be simply estimated through 
\begin{align}
\overline{\tau} = -2E\left\langle \frac{\mathrm{d}E}{\mathrm{d}t}
\right\rangle^{-1}\,,\label{eq:damping_tau}
\end{align}
where $\left\langle\cdot\right\rangle$ denotes the time-average over one
pulsation period. It is interesting to note that a similar formula can be
derived from the simple damped harmonic oscillator by assuming that
$\overline{\tau}\gg 2\pi \omega^{-1}$, where $\omega$ is the angular
frequency of the corresponding undamped case.

Truncating $E$ and $\mathrm{d}E/\mathrm{d}t$ at leading order in the
post-Newtonian expansion, see~\cite{CerdaDuran2010} for details, we
obtain the following expressions
\begin{align}
E&=4\pi
\int_0^R\mathrm{d}rr^2\sqrt{-g}\frac{1}{2}\rho_{\mathrm{_B}}v'^2\,,\\
\frac{\mathrm{d}E}{\mathrm{d}t}&=-4\pi\int_0^R\mathrm{d}rr^2\sqrt{-g}\zeta|\Theta|^2\,,
\end{align}
where $v'=v-v_{\mathrm{_B}}=v$ is the Newtonian three-velocity perturbation in the
radial direction with $v_{\mathrm{_B}}$ being the background velocity.
The background velocity is set to zero
in our case, \ie $v_{\mathrm{_B}}=0$. Analogously, $\rho_{\mathrm{_B}}$ is the background density profile of
the TOV solution. 

Assuming a harmonic form of the perturbation, \ie $v'\propto \exp[i\omega
t+ikr]$, where $k$ is the wavenumber, we obtain 
\begin{align}
\left\langle |\Theta|^2 \right\rangle\propto k^2\left\langle
v'^2\right\rangle = k^2v_{\mathrm{max}}'^2/2\,. 
\end{align}
The quantity $v_{\mathrm{max}}'$ is a function of radius. Analogously to
the simple harmonic oscillator, the energy oscillates between
kinetic and potential energy. Therefore $E$ can be estimated by
substituting $v_{\mathrm{max}}'$ for $v'$. Then, by using the definition of
the speed of sound $c_s^2=\omega^2/k^2$ and assuming $\zeta$ to be
constant within the star we obtain
\begin{align}
E&=2\pi
\int_0^R\mathrm{d}rr^2\sqrt{-g}\rho_{\mathrm{_B}}v_{\mathrm{max}}'^2\,,\\
\left\langle\frac{\mathrm{d}E}{\mathrm{d}t}\right\rangle&\sim-2\pi
\frac{\omega^2}{c_s^2}\zeta\int_0^R\mathrm{d}rr^2\sqrt{-g}v_{\mathrm{max}}'^2\,.
\end{align}

Finally, by using Eq.~\eqref{eq:damping_tau} we find a formula which
relates the damping time of radial pulsations $\overline{\tau}$ to the
bulk viscosity $\zeta$
\begin{align}
\zeta =
\frac{2}{\overline{\tau}}\frac{c_s^2}{\omega^2}\overline{\rho}\,,
\label{eq:visfromtau}
\end{align}
where
\begin{align}
\label{eq:rho_av}
\overline{\rho}:=\frac{\int_0^R\mathrm{d}rr^2\sqrt{-g}v_{\mathrm{max}}'^2\rho_{\mathrm{_B}}}{\int_0^R\mathrm{d}rr^2\sqrt{-g}v_{\mathrm{max}}'^2}\,,
\end{align}
is the eigenmode-averaged background density. In practice, $\omega$ can
be obtained from the simulations by measuring the pulsation frequency,
$\overline{\rho}$ can be obtained from the initial data by calculating
the integrals in Eq.~\eqref{eq:rho_av} and $c_s^2$ can be simply
evaluated at $\overline{\rho}$ by using the polytropic part of the
EOS, \ie
\begin{align}
c_s^2 = \left[\frac{1}{\Gamma \kappa {\overline{\rho}}^{\Gamma -1}}+
\frac{1}{\Gamma-1}\right]^{-1}\,.
\end{align}
This means that we are left with the determination of
$\overline{\tau}$ in order to calculate $\zeta$ through
Eq.~\eqref{eq:visfromtau}.

As already mentioned above, we use the central density
$\rho_{c}$ to
measure the damping time $\overline{\tau}$. In order to do so, we make use of the
following simple recipe:

\begin{itemize}
\item[1.]\textit{Use a high-order low-pass Butterworth filter in order to remove
global drifts of the central density from the signal.}

As can be seen from the left panel of Fig.~\ref{fig:osc_evol}, our simulations have a
typical length on the order of $\sim\mathcal{O}(100~\mathrm{ms})$ which
yields a minimum resolved frequency of $f_{\mathrm{min}}\sim
0.01~\mathrm{kHz}$. We choose the cutoff-frequency on the order of
$f_c\sim 0.1~\mathrm{kHz}$ and a $n=4$ filter, where $n$ denotes the
order of the Butterworth filter. The frequency of the fundamental mode is
$f_F\sim 1.44~\mathrm{kHz}$, see \eg~\cite{Font02c}, which means that
our cutoff-frequency is more than an order of magnitude lower than the
frequency we want to resolve. Thus, the choices $f_c\sim
0.01~\mathrm{kHz}$ and $n=4$ ensure that we obtain the density drift
without contributions from higher frequencies. Then, we subtract the
drift from the original data and obtain a signal which oscillates around
zero. Afterwards, we normalize the signal by the initial
central density $\rho_{c,0}$. We define the ``drift-free'' and
normalized signal as $\delta\rho$ which is shown in the left panel of
Fig.~\ref{fig:osc_evol} in solid thin lines for the simulations with the
highest resolution.

\item[2.]\textit{Calculate the local maxima of the signal.}

Assuming that the signal is composed of a sum of damped sinusoids, we can
calculate the local maxima of the signal in order to obtain data which
can be fitted to a simple exponential function. The main caveat in this
approach is that it is not possible to distinguish between different
excitation modes with different damping times. This means that it is
necessary select only those parts of the timeseries of $\delta \rho$
which are dominated by a single eigenmode; the fundamental or $F$-mode in
our case. This is not a problem for lower resolution simulations as the
overtone modes have a shorter damping time due to more efficient
numerical damping. 

Therefore, for lower resolution simulations the contamination of the signal
is reduced such that almost all of $\delta \rho$ is dominated by the
$F$-mode. However, as numerical damping decreases with increasing
resolution also the contributions from excited overtone modes become
non-negligible. This can be seen in the left panel of Fig.~\ref{fig:osc_evol} where for $t
\lesssim 50~\mathrm{ms}$ the evolution of the local maxima is highly
oscillatory due to contributions from overtone modes. A naive computation
of the local maxima together with a subsequent fitting procedure results
in an underestimation of $\overline{\tau}_F$ because overtone modes have
shorter damping times than the fundamental mode. Thus, for
higher resolution simulations we make only use of approximately the last
half of the signal where the contributions from the overtone modes can be
neglected. We show thick solid lines which connect the local maxima in
the left panel of Fig.~\ref{fig:osc_evol} in order to visualize the utilised part of
$\delta \rho$. The timeseries of the utilised local maxima of $\delta
\rho$ is defined as $\widehat{\delta \rho}$ and the first value in this
sequence is defined as $\widehat{\delta \rho}_0$.

\item[3.]\textit{Fit the logarithm of the local maxima to a linear function in
time.}

In this step we additionally normalize $\widehat{\delta \rho}$ by
$\widehat{\delta \rho}_0$ and fit the values of $\log[\widehat{\delta
\rho}/\widehat{\delta \rho}_0]$ to a linear function in time. The results
are shown in the left and right panels of Fig.~\ref{fig:osc_evol}. Dashed
lines in the left panel of Fig.~\ref{fig:osc_evol} and the star symbols
in right panel of Fig.~\ref{fig:osc_evol} represent the fits,
respectively. To avoid overcrowding the right panel of
Fig.~\ref{fig:osc_evol}, we only show $\widehat{\delta
\rho}/\widehat{\delta \rho}_0$ for the high-viscosity case. Different
color intensities in the right panel of Fig.~\ref{fig:osc_evol} show
different resolutions and solid lines connect the original data points of
$\widehat{\delta \rho}/\widehat{\delta \rho}_0$. 
\end{itemize}

From the right panel of Fig.~\ref{fig:osc_evol} we observe convergent
behaviour in the slopes of the linear fits. This is expected because
increased numerical resolution leads to a decrease of the numerical
viscosity such that the measured damping time converges to the damping
time related to the physical viscosity. Finally, we observe oscillatory
behaviour in the data presented for the highest resolution in the right
panel of Fig.~\ref{fig:osc_evol}. As already discussed, these
oscillations stem from contributions of overtone modes and do not
significantly impact the calculation of $\overline{\tau}_F$.

\begin{figure}
\includegraphics[width=0.49\textwidth]{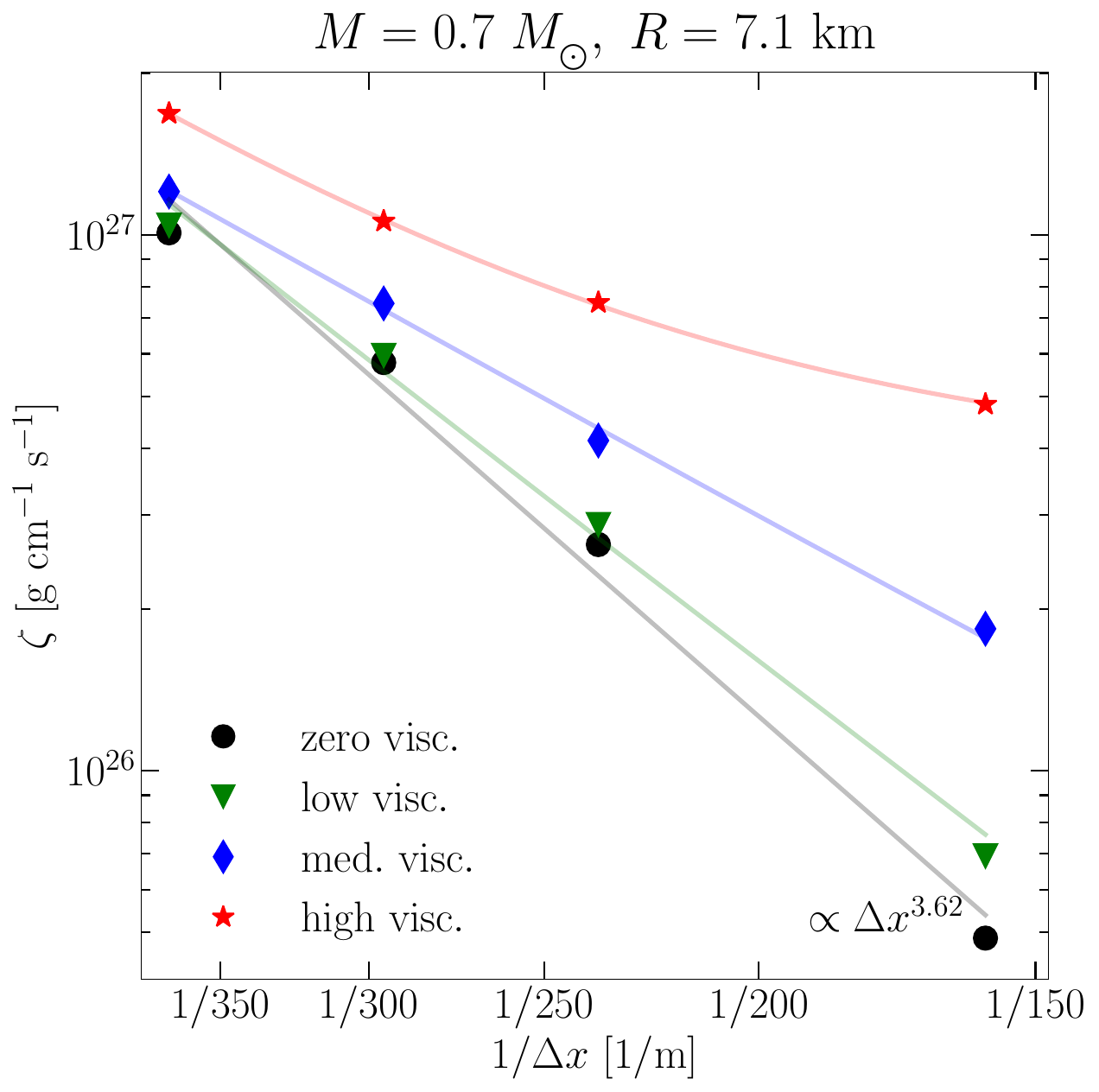}
\caption{Same as Fig.~\ref{fig:num_vis} but for a
TOV solution with a smaller mass and radius.
\label{fig:num_vis_old}
}
\end{figure}

\section{Convergence}
\label{sec:small_star}

Because of the reduced convergence order found in Fig.~\ref{fig:num_vis}
it is interesting to compare to another numerical viscosity measurement
performed during an earlier development stage of our bulk-viscous
extension of \texttt{FIL}. In that case, we employed a different TOV
solution where $\kappa=25$, $\Gamma_{\mathrm{th}}=2$ and
$\rho_c=5.12\times10^{-3}~M_{\odot}^{-2}\approx 3.16\times
10^{15}~\mathrm{g}~\mathrm{cm}^{-3}$. All other parameters of the EOS are
the same. This setup yields a $M=0.7~M_{\odot}$ star with a
$R=7.1~\mathrm{km}$ radius. The bulk viscosity is varied between
$\zeta_{h} \in [0, \sim 2.73 \times 10^{25}, \sim 1.64 \times 10^{26},
  \sim 5.47 \times 10^{26}]~\mathrm{g}~\mathrm{s}^{-1}~\mathrm{cm}^{-1}$,
which are denoted as zero, low, medium and high-viscosity cases,
respectively. The resolution on the finest refinement level varies
between $\Delta x \in \{\sim 158, \sim 236, \sim 295, \sim
369\}~\mathrm{m}$. Furthermore, instead of employing the cubic polynomial
presented in Eq.~\eqref{eq:zeta_bound}, $\zeta$ is set to zero sharply
for $\rho < 2.56 \times 10^{-4}~M_{\odot}^{-2}\approx 1.58 \times
10^{14}~\mathrm{g}~\mathrm{cm}^{-3}$.

Figure~\ref{fig:num_vis_old} presents the same numerical viscosity
measurement as shown in Fig.~\ref{fig:num_vis} for the small TOV
solution. Table~\ref{tab:fits_old} presents the corresponding fitting
coefficients. First, we observe that only for the high-viscosity case
the fit shows a concave shape indicating that the low and
medium-viscosity cases are dominated by numerical viscosity, even at high
resolutions. More importantly, we also find that the convergence order
for the zero viscosity case,\ie $\sim 3.62$, is consistent with the
formal convergence order of our numerical schemes. This is in contrast to
the reduced convergence order of $\sim 1.79$ found in
Fig.~\ref{fig:num_vis}. We suspect that this behaviour is related to the
amount of grid points covering the stars in both cases. For the highest
resolution of the setup presented in Fig.~\ref{fig:num_vis_old}, the
neutron star is covered by only $\sim 66\%$ of the points of the more
realistic bigger star simulated at the highest resolution in
Fig.~\ref{fig:num_vis}. This is a direct consequence of the smaller
radius of the setup in Fig.~\ref{fig:num_vis_old} which results in a
lower effective resolution.

Now, the reduced effective resolution of the setup in
Fig.~\ref{fig:num_vis_old} increases the numerical error originating from
the interior of star. Due to the smooth behaviour of the solution in the
neutron star interior, the convergence order in this region is expected
to be close to the formal one. Hence, we suspect that the setup in
Fig.~\ref{fig:num_vis_old} leads to a measured convergence order closer
to the formal one because the damping of central density oscillations is
dominated by discretization errors which originate from the low effective
resolution in the neutron star interior. In contrast, the damping of the
zero-viscosity case in Fig.~\ref{fig:num_vis} is dominated by
discretization errors from the ill-balanced neutron star surface which
leads to a deterioration of the convergence order. This is a direct
consequence of the larger radius which leads to small discretization
errors from the neutron star interior due to a higher effective
resolution.

\begin{table}
\centering
\begin{tabular}{lcccc}
\hline
\hline
Model
& $\zeta_a$ & $\zeta_s$ & $p_h$ \\
\hline
& $[\mathrm{g}~\mathrm{s}^{-1}~\mathrm{cm}^{-1}]$ & 
$[\mathrm{g}~\mathrm{s}^{-1}~\mathrm{cm}^{-1}]$ & 
 \\
\hline
zero visc. & $--$ & $1.66 \times 10^{33}$ & $3.62$ \\[3pt] 
low visc.  & $--$ & $3.49 \times 10^{32}$ & $3.23$ \\[3pt]
med. visc. & $8.87\times 10^{24}$ & $1.04 \times 10^{31}$ & $2.32$ \\[3pt]
high visc. & $3.66\times 10^{26}$ & $8.29 \times 10^{31}$ & $2.83$ \\
\hline
\hline
\end{tabular}
\caption{Fitting results to the measured bulk-viscosity values
    presented in Fig.~\ref{fig:num_vis_old} using
    Eq.~\eqref{eq:numvisfit} with $\lambda \approx 1.74 \times
    10^{4}~\mathrm{m}$. Note that the values of $\zeta_a$ for the zero
    and low-viscosity case are not available due to the fact that we
    applied a linear fit of $\log[\zeta]$ as a function of $\log[\Delta
      x]$ for both cases. Using Eq.~\eqref{eq:numvisfit} for the
    low-viscosity case resulted in negative values for $\zeta_a$ due to
    the convex shape of the corresponding data points in
    Fig.~\ref{fig:num_vis_old}.}
\label{tab:fits_old} 
\end{table}

\section{Rotational properties of neutron star mergers with a constant
bulk viscosity}
\label{sec:bnsrot}

\begin{figure*}
\includegraphics[width=0.33\textwidth]{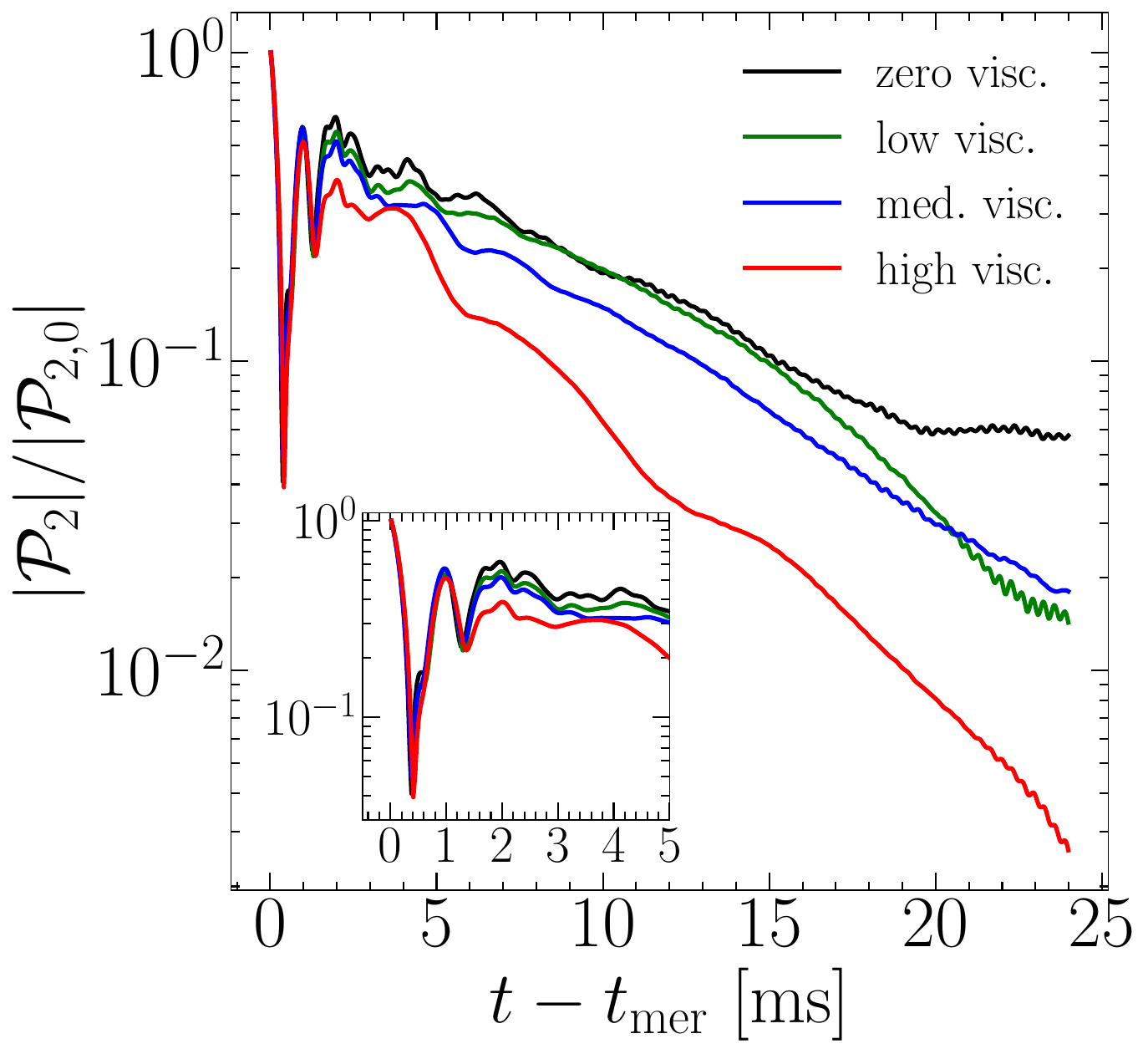}
\hskip 0.1cm
\includegraphics[width=0.33\textwidth]{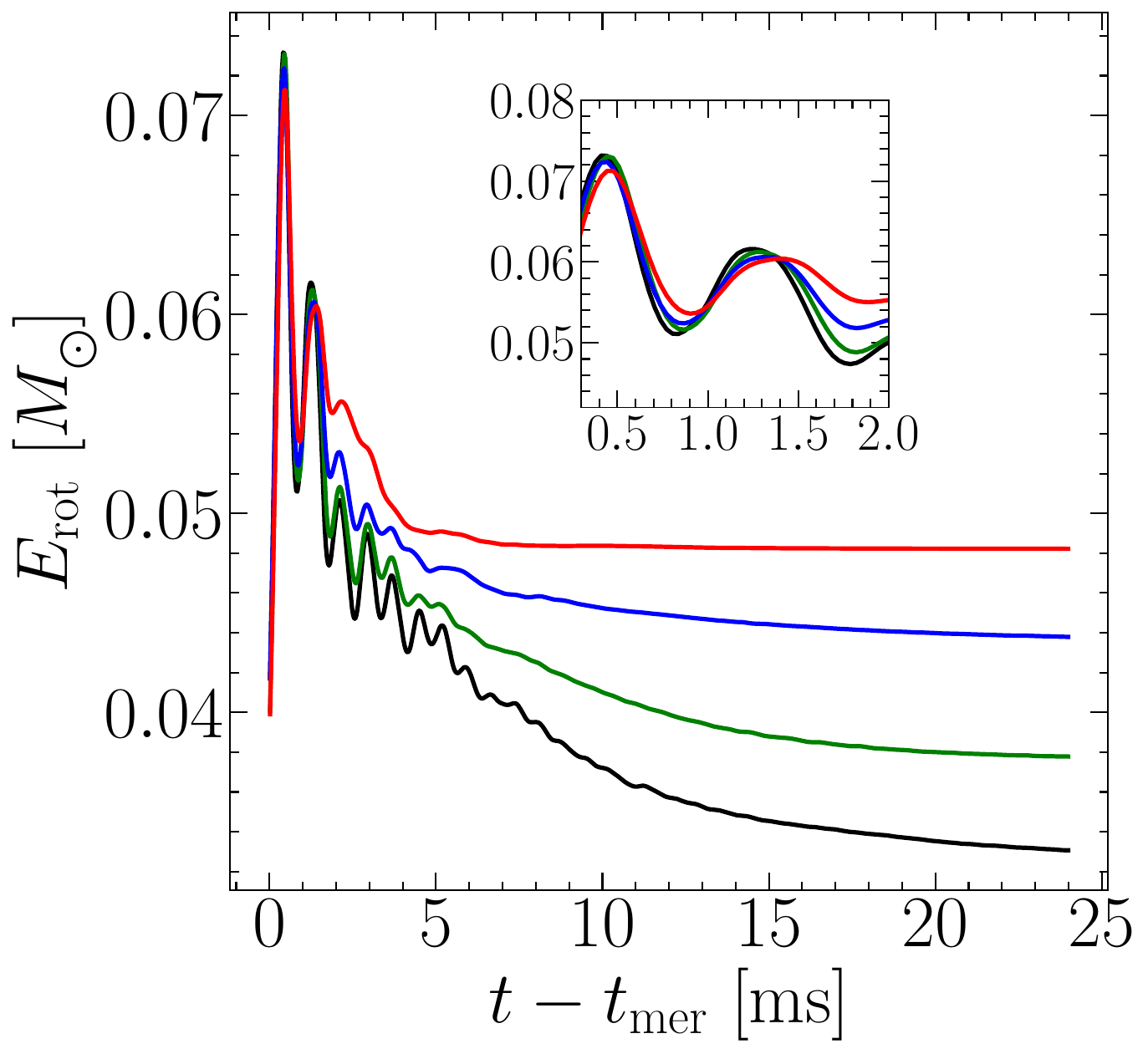}
\hskip 0.1cm
\includegraphics[width=0.31\textwidth]{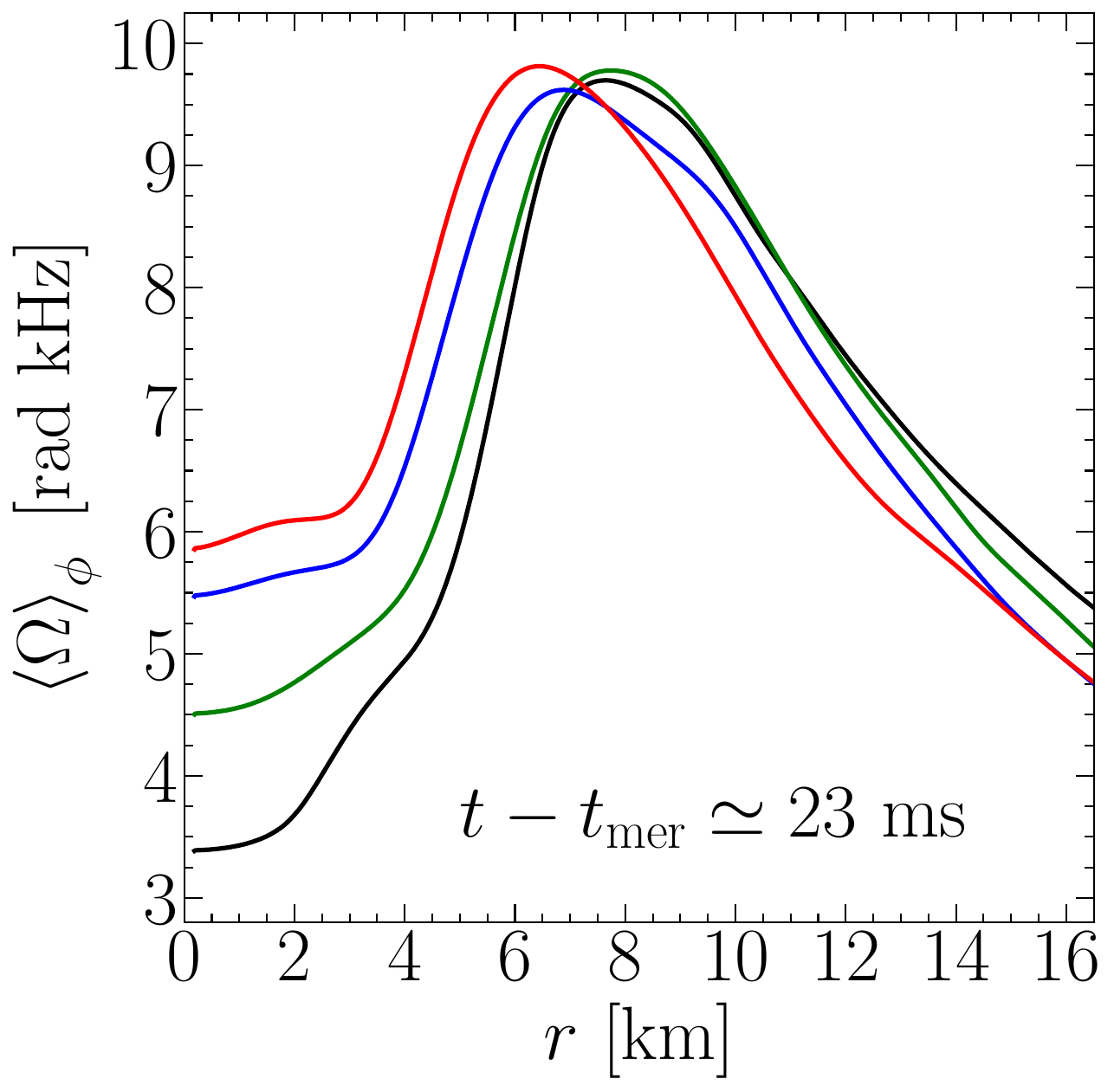}
\caption{\textit{Left:} Evolution of the $m=2$ rest-mass density mode
  $\mathcal{P}_{2}$ normalized to its value at the time of the merger
  $\mathcal{P}_{2,0}$ for the four constant viscosity cases
  considered. \textit{Middle:} The same as on the left but for the
  rotational kinetic energy. \textit{Right:} $\phi$-averages in the 
  equatorial plane of the
  angular velocity as a function of the coordinate radius $r$ at a
  representative late times ($t-t_{\mathrm{mer}}\approx
  23\,\mathrm{ms}$).}
  \label{fig:mtwo}
\end{figure*}

\begin{figure}
\includegraphics[width=0.49\textwidth]{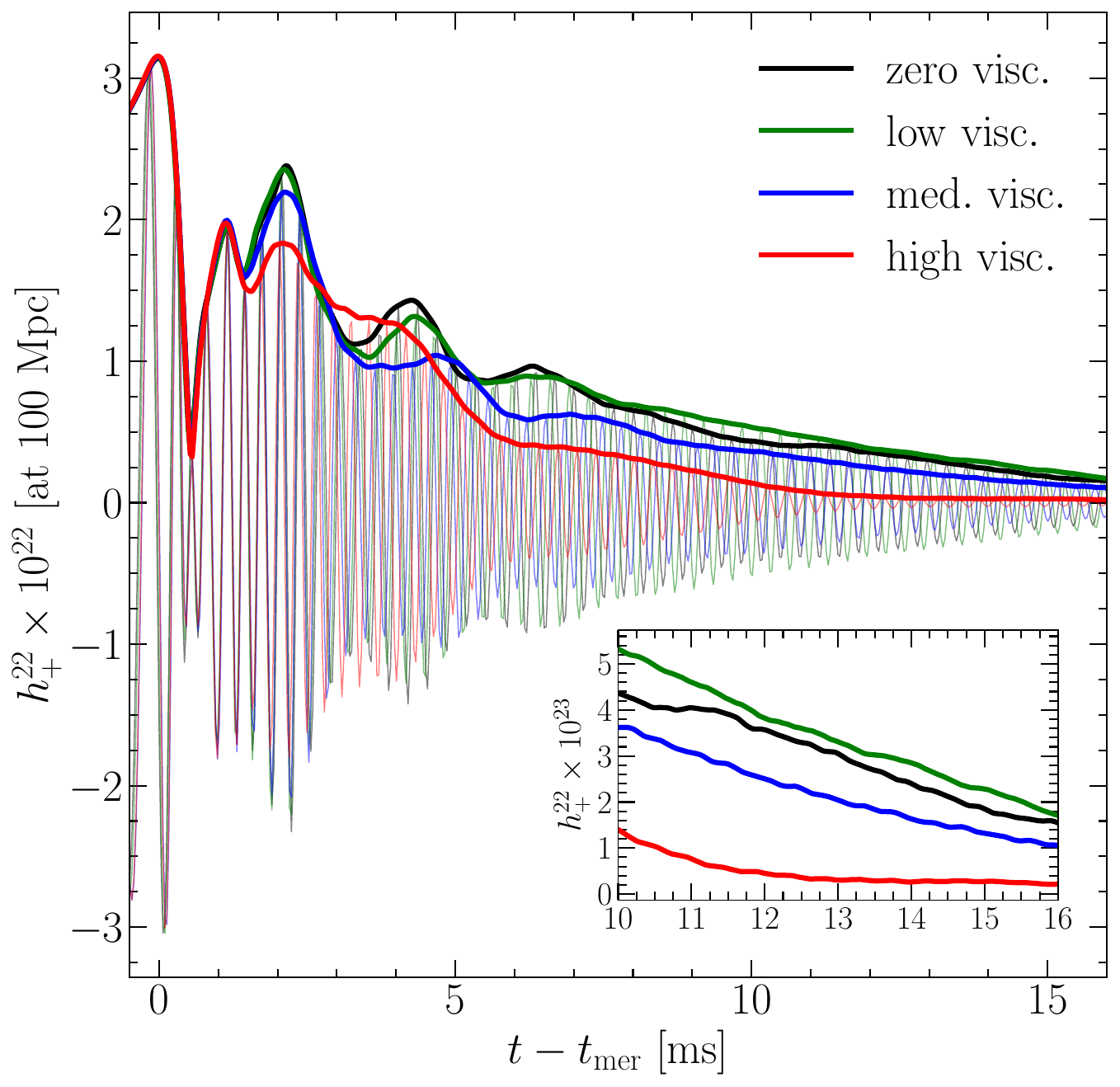}
\caption{GW strain in the $\ell=m=2$ mode of the $+~$--polarization
    extracted at $\sim 740\,\mathrm{km}$ and normalized to a distance of
    $100\,{\rm Mpc}$ for the four configurations considered. Thick solid
    lines report the corresponding amplitudes.}
\label{fig:waves}
\end{figure}

The left panel of Fig.~\ref{fig:mtwo} reports the evolution of the $m=2$
density mode, or bar-mode~\cite{Baiotti06b}
\begin{align}
\mathcal{P}_{m} :=\int \rho W e^{-im\phi}\, \sqrt{\gamma}\, dxdydz\,,
\qquad m=2\,,
\label{eq:mode}
\end{align}
when normalized to its value at the time of the merger
$\mathcal{P}_{2,0}$, and for the four values of constant bulk viscosity
considered in our simulations, see also \cite{Chabanov2023}. Note that
the initial oscillations for $t-t_{\rm mer} \lesssim 3\,{\rm ms}$ are
the result of the rapid and quasi-periodic collisions of the two
stellar cores (see~\cite{Takami2015} for a mechanical toy
model). Furthermore, the evolution of $\mathcal{P}_{2} /
\mathcal{P}_{2,0}$ is impacted in a systematic manner by the strength
of the bulk viscosity, which suppresses the $m=2$ deformation and hence
leads to a more axisymmetric HMNS.
Note that when moving away from the constant value adopted in this
work, a microphysical model for the bulk-viscosity leads to increased
bar-deformations \cite{Chabanov2023}. Since the bar-deformation
represents a way in which the HMNS minimises its rotational
energy~\cite{Baiotti06b}, it is natural to expect that bulk viscosity
will equally impact the rotational properties of the HMNS. We quantify
the latter by computing the rotational kinetic energy of the merger
remnant [see, \eg Eq.~(12.48) of~\cite{Rezzolla_book:2013} for a
definition].

The middle panel of Fig.~\ref{fig:mtwo} reports the evolution of the
rotational energy. We observe that the late-time rotational energy
depends on the strength of the bulk viscosity, with larger bulk
viscosities leading to a higher $E_{\mathrm{rot}}$\footnote{A
  non-monotonic behaviour is observed for the late-time evolution of
  $\mathcal{P}_{2}$ in the cases of $\zeta=0.2\zeta_0$ and
  $\zeta=0.5\zeta_0$; this is not particularly surprising given that
  the differences in viscosity are comparable with the fluctuations
  introduced by the turbulent motion of the fluid (see also
  \cite{Radice2017} for a similar behaviour.}. To understand this
result it is sufficient to bear in mind that by removing
non-axisymmetric deformations, bulk viscosity is effectively quenching
GW emission and hence conserving the (rotational) energy of the HMNS.
 
Further evidence for this dynamics is provided in the right panel of
Fig.~\ref{fig:mtwo}, which shows the azimuthal ($\phi$) averages in the
equatorial plane of the angular velocity as a function of the coordinate
radius $r$ at a representative late time, \ie $t-t_{\mathrm{mer}} \approx
23\,\mathrm{ms}$ (time averages over $2\,$ ms yield very similar
profiles). While this is appreciable already early in the post-merger
evolution, \ie at $t-t_{\mathrm{mer}} \lesssim 5\,\mathrm{ms}$, the right
panel of Fig.~\ref{fig:mtwo} clearly shows that the angular velocity of
the HMNS core (left panel), \ie for $r\lesssim 8\,\mathrm{km}$ is larger
in the case of large bulk viscosities and after reaching a local maximum
at $\simeq 4-5\,{\rm km}$, it falls off following a Keplerian profile. In
turn, the larger centrifugal and thermal support available in the central
regions of the HMNS with large viscosities changes the stellar structure
and leads to progressively smaller values of the rest-mass density in the
core of the HMNS, see left panel of Fig.~\ref{fig:vis_thermal}, and to
larger values in the envelope (not shown); as a result, the peak of the
angular velocity moves inwards with increasing bulk viscosity.

Overall, by comparing our simulations in this work, which employ a
constant bulk viscosity within the neutron stars, with those utilizing a
model based on realistic microphysical arguments \cite{Chabanov2023} we
observe an ``inverted'' behaviour of $\mathcal{P}_{2}$ as described
above. As a result, GW emission is decreased instead of increased such
that the rotational properties of the HMNS remnant are altered towards
larger rotational energies and less compact remnants. We support this
point of view by reporting the post-merger GW signal in
Fig.~\ref{fig:waves}.

\bibliography{aeireferences.bib}

\end{document}